\documentclass[aps,prd,twocolumn,groupedaddress,nofootinbib,showkeys]%
{revtex4-1}

\pdfoutput=1

\usepackage{comment}

\usepackage[T1]{fontenc}
\usepackage{enumerate}
\usepackage{bm}
\usepackage{amsmath}
\usepackage{amssymb}
\usepackage{graphicx}
\newcommand{\R}{\Bbb R}
\newcommand{\C}{\Bbb C}
\newcommand{\T}{\Bbb T}
\newcommand{\MM}{\Bbb M}
\newcommand{\Spin}{\Bbb S}
\newcommand{\thorn}{\text{\TH}}
\newcommand{\omicron}{{o}}
\newcommand{\RR}{{\mathcal{R}}}
\newcommand{\TT}{{\mathcal{T}}}

\newcommand{\scrif}{{\mathcal{I}^{+}}}

\newcommand{\const}{\mathrm{const}}
\newcommand{\conj}{\mathrm{conjugate}}

\newcommand{\z}{{\mathfrak z}}

\newlength{\slashh}
\newcommand\fanm{\settoheight{\slashh}{$/$}$/\!\raisebox{.5\slashh}{$\smile$}\!\backslash$}

\begin{document}

\title{Angular momentum, spinors and twistors}

\author{Adam D. Helfer}

\email[]{helfera@missouri.edu}
\affiliation{Department of Mathematics and Department of Physics \& Astronomy,
University of Missouri,
Columbia, MO 65211, U.S.A.}

\date{\today}

\begin{abstract}
Twistors appear to provide a satisfactory treatment of angular momentum for gravitationally radiating systems.  The approach is manifestly Bondi-Metzner-Sachs (BMS) invariant, and there are no supertranslation ambiguities.  
The resulting definitions of center of mass and spin are appealing:  unphysical contributions from bad cuts are canceled off from the center of mass, and the spin appears formally as a displacement of the center of mass into the complex.  
For transitions between asymptotically Minkowskian regimes (non-radiative regimes with purely electric Bondi shear), when there is no supertranslation offset (equivalently, radation memory) between the regimes the results are in agreement with those deducible from other approaches.  However, when there is an offset, the results are different.  
The twistor-derived change-of-origin formula is closely parallel to the special-relativistic one, 
with an algebraic cross-product between the energy--momentum and a direction-dependent translation derived from the supertranslation.  (No supermomenta appear.) 
There is also a ``longitudinal'' contribution to the emitted angular momentum
(one sensitive to the total energy--momentum and not just the emitted energy--momentum), and terms which are both linear and cubic in the gravitational radiation (whereas BMS-based definitions give purely quadratic contributions).
The first-order terms mean that the supertranslation offsets can contribute to the exchange of angular momentum with gravitational radiation, even in weak-field limits.
This is illustrated with a
simple almost-special-relativistic model.
\end{abstract}


\maketitle

\section{Introduction}

Energy--momentum and angular momentum are fundamental quantities in non-general-relativistic physics, and one would like to extend them to general relativity.  We may hope an extension will give insights to what the most important kinematic degrees of freedom are, and even guides for how to reconcile general relativity with quantum theory.

An important class of systems to consider are those which are
isolated and emit gravitational radiation --- Bondi--Sachs space--times \cite{Bondi1960,Sachs1962a,Sachs1962b}.  A good understanding of these would mean not just identifying the total energy--momentum or angular momentum, but quantifying what is carried off as radiation is emitted.

Bondi and Sachs, in the first papers, put forward a definition of energy--momentum.  Evidence for it has accumulated, and it is now broadly accepted.  Angular momentum, though, has been more problematic.  In fact several, disparate, approaches have been suggested; these differ in what sort of quantity the angular momentum is taken to be and how it might be applied (see Ref. \cite{Szabados2009} for a review).  Those discrepancies reflect unsettled questions about the foundations of the subject.

The primary purpose of this paper is to give a new treatment of one of those approaches, the twistor one \cite{Penrose1982,PR1986,ADH2007,ADH2009}.  This proposal so far seems satisfactory, being manifestly Bondi--Metzner--Sachs invariant, having no supertranslation ambiguity, and providing attractive definitions of spin and center of mass.\footnote{Relativistically, one should, strictly speaking, call this the center of energy, and the corresponding vector the first energy moment, but I will keep the conventional terminology.}  The treatment here, while explaining the twistor motivations, is cast in conventional space--time terms.  No prior knowledge of twistor theory is needed.

To explain the ideas and results more fully, and especially to clarify the motivations and consequences of the fundamentally different choices involved in different approaches, I will begin with some background.

\subsection{Null infinity and Bondi shear}

For Bondi--Sachs space--times, Penrose's future null infinity $\scrif$ has certain universal properties \cite{PR1986}.  
It is naturally a bundle of affine lines over $S^2$, and we coordinatize
$\scrif\simeq \R\times S^2$ as $(u,\theta ,\phi )$, with $u$ a {\em Bondi retarded time parameter} and $(\theta ,\phi )$ spherical polars.  Here the sphere may be identified with the space of asymptotic null directions, and each fiber is called a {\em generator} of $\scrif$.  The group preserving the universal structure is the {\em Bondi--Metzner--Sachs (BMS) group,} and it is the semidirect product of the proper orthochronous Lorentz group (which acts naturally on $S^2$) and the {\em supertranslations,} of the form $u\mapsto \acute{u} =u+\alpha (\theta ,\phi )$ for suitably smooth $\alpha$.  In particular, all {\em cuts} of $\scrif$, that is, cross-sections $u=\z (\theta ,\phi )$ for suitably smooth $\z$, are on equal footing, as far as the universal structure goes.

A cut $u=\z (\theta ,\phi)$ 
can be thought of as an
instant of retarded time (demarking radiation already emitted, in the regime $u\leq \z$, from degrees of freedom remaining within the system).  
We thus seek the energy--momenta and angular momenta at arbitrary cuts of $\scrif$.  This is much more demanding than asking for the total energy--momentum or angular momentum. 

For any Bondi--Sachs space--time, a
key quantity is the {\em Bondi shear} $\sigma$; it has dimension length.  It is usually said to be a (spin-weight two) function on $\scrif$, but that is not quite accurate; it depends also on the choice of Bondi coordinates (so it is a functional of that choice, and, given that choice, a function on $\scrif$).  Under a passive supertranslation as above, it changes as $\acute\sigma =\sigma -\eth^2\alpha$, where $\eth$ is an antiholomorphic derivative in the angular directions.  This can be regarded as a gauge change, and it will be important shortly.  The shear of a cut is the value of the shear in a Bondi system for which the cut is a constant value of the coordinate $u$; a cut is called {\em good} if it is shear-free, and {\em bad} otherwise.

The shear admits an angular potential $\lambda$, such that $\sigma =\eth^2\lambda$.
In general, the potential $\lambda$ must be taken to be complex, and the parts $\eth^2\Re\lambda$, $i\eth^2\Im\lambda$ are called the {\em electric} and {\em magnetic} parts of the shear.  

Notice that the supertranslational freedom only affects the electric part of the shear.
(A magnetic contribution to the shear is thus always an obstruction to finding a good cut, and to identifying a cut with one for Minkowski space.)  It is also true that the supertranslations which are identifiable as asymptotic translations are those with $\eth^2\alpha=0$, so translations do not alter the shear.  (While this gives an invariant identification of the translations as a subgroup of the supertranslations, there is no invariant sense of a ``pure,'' that is, translation-free, supertranslation.)

By a {\em regime} on $\scrif$, I will mean the open region bounded by two disjoint cuts (or the semi-infinite region to the future or past of one cut, or all of $\scrif$).  

The quantity $\dot\sigma$ (where the overdot denotes $\partial/\partial u$) signals the presence of gravitational radiation.  A non-radiative regime thus has $\dot\sigma =0$.  Non-radiative regimes with magnetic shear are in principle possible, but are usually considered exotic, and it is suspected that  in realistic situations they can at exist only transiently.
Therefore most work assumes that the non-radiating regimes of generic interest have $u$-independent purely electric shear.  I will call those {\em Minkowskian,} for reasons which will become apparent.\footnote{A stationary regime is necessarily Minkowskian, but stationarity is too strong a condition to encompass many situations of interest.  For example, if a system fissions the corresponding regime may well become Minkowskian but not stationary.}

If all of $\scrif$ is a Minkowskian regime, then one can solve the equation $\sigma =\eth^2\alpha$ for a supertranslation $\alpha$, and there will be a four-dimensional family of good cuts, those which are translations of $\alpha$.  In fact, this family can naturally be given the structure of Minkowski space, and virtually all workers accept that, in such a situation, these cuts should be interpretable as origins for angular momentum.  (Had we started from Minkowski space, the good cuts would be the intersections of the future null cones of points with $\scrif$.)  
If a finite Minkowskian regime persists long enough, it will also admit a four-dimensional family of good cuts, interpretable as an open set in a Minkowski space, and these would be accepted as origins for the definition of angular momentum in the regime.\footnote{We will see below that there are Minkowskian regimes which do not persist long enough to admit good cuts, but we will also see that in the twistor approach this distinction is not significant.}
However, if there are two distinct Minkowskian regimes, their families of good cuts will generally be relatively supertranslated, that is, there is a {\em supertranslation mismatch} or {\em offset}. 
(When the two offset regimes encompass the $u\to\pm\infty$ parts of $\scrif$, one sometimes says the shear lies in an {\em infrared sector}.)
This means that there can be no invariant Poincar\'e motion identifying the corresponding Minkowski spaces; this is the obstruction to finding a space of origins modeled on Minkowski space.  It is tied to gravitational radiation memory.

\subsection{Asymptotic covectors}

Bondi and Sachs gave a definition of the energy--momentum $\bm{P_a}(\z )$ at
any cut, which is broadly accepted and powerful.  
Its successes, which involve the formula itself, have perhaps distracted from another, highly non-trivial, aspect of the construction:  the identification of 
a single
asymptotic cotangent space (call it $T^*_\scrif\simeq\R ^4$, independent of $\z$) in which $\bm{P_a}(\z )$ takes values \vphantom{$\bigl[$}\cite{ADH2014}.  This space is rather indirectly (and, in the original papers, rather implicitly) defined.
It is {\em not} a space of asymptotically covariantly constant covector fields, for example --- there {\em is} no such space, when gravitational radiation is present.  That it is nevertheless possible to identify $T^*_\scrif$ is a consequence of the depth of the  Bondi--Sachs structure.
It is the fact that the energy--momenta for different cuts $\z$ all take values in this one space which enables us to compare them, and to say how much is carried off in radiation.

Indeed, the space $T^*_\scrif$ is not just cut-independent, but {\em universal,} that is, it depends only on the structure common to all Bondi--Sachs space--times.  This means that energy--momenta of different space--times can be compared (once we fix a BMS motion identifying their $\scrif$'s), just as in special relativity we can compare the energy--momenta and angular momenta of two independently-given systems (once we fix a Poincar\'e motion specifying the relation between the two frames of reference).  It is universality we tacitly appeal to whenever we think of the energy--momentum as being significant, not just for each Bondi--Sachs space--time individually, but as a quantity of interest in comparing different systems.

\subsection{Origins and angular momentum tensors}

For angular momentum, the situation is more delicate, due to its origin-dependence.
I will show below that the precise way this 
is encoded is critical, and that when twistor theory is used a transition to general relativity is not problematic.  But 
I will start with a conventional treatment.

Conventionally in
special relativity, angular momentum is a tensor field $M_{ab}(x)$, where $x$ is the origin about which it is measured.
If we want a similar formulation in general relativity, then we must go beyond defining a {\em vector space} $T^*_\scrif$ (or its dual $T_\scrif$), and come up with some sort of {\em affine space} to serve as the set of origins.  But the integrability conditions for such a set are even more stringent than those for $T^*_\scrif$.

These difficulties appear immediately.  As noted above, in a Minkowskian regime, it is generally accepted that the four-dimensional family of shear-free cuts should be identifiable with the set of origins, but two such regimes will generally be relatively supertranslated.
This means that there can be no Poincar\'e motion satisfactorily identifying the spaces of origins for the two regimes.  The obstruction to integrability is explicitly calculable as the gravitational radiation memory \cite{ADH2007}.

This is a significantly negative finding, and it bears some reflection.  Differing responses to it have resulted in very different approaches to angular momentum.  Some workers (notably Newman and followers --- see \cite{Newman1976,Kozameh_2005,Kozameh_2008,Adamo_2009,Kozameh_2020} and references therein), in effect, 
drop the requirement that the space of origins be affine.  Others drop the four-dimensionality, and indeed it is most common nowadays to pass to infinite-dimensional spaces of origins (as in the BMS-based approaches of Ashtekar and Streubel \cite{AS1981,ADH2020}, and of Dray and Streubel \cite{DS1984,Dray1985}; the idea using the BMS group goes back to Winicour and Tamburino \cite{WT1965}).

\subsubsection{BMS-based approaches}

The BMS-based approaches exploit the formal parallel with the Poincar\'e group.  To develop this, they take the space of {\em all} cuts as the space of origins.  The angular momentum then becomes a function $\bm{M_{ab}}(\z_{\rm act}, \z_{\rm pas})$ of two arbitrary cuts, an {\em active} one $\z_{\rm act}$ specifying the instant of retarded time at which we wish to evaluate the kinematics, and a {\em passive} on $\z_{\rm pas}$ representing the choice of origin.\footnote{This terminology is due to Szabados \cite{Szabados2009}.  It does not correspond to the usual senses of active and passive transformations, but it does make a distinction which is essential here.  For a conventional special-relativistic $M_{ab}(x)$, there would be no active dependence (because it is a total angular momentum); the choice of $x$ would be the passive dependence.}
This means that even at a fixed active cut the angular momentum is an infinite-dimensional object, and indeed one must bring in an infinite-dimensional family of new kinematic quantities, the {\em supermomenta,} to treat the origin-dependence.

The resulting structure is a mathematically well-defined system of ``BMS charges,'' dual to the generators of the BMS group, modeled on the duality between the Killing vectors in Minkowski space and the energy--momentum and angular momentum.  
We do need more work
to be confident that the charges $\bm{M_{ab}}(\z_{\rm act}, \z_{\rm pas})$ which are {\em called} angular momenta are being interpreted properly.  
(The ``obvious'' definitions of spin and center of mass associated with $\bm{M_{ab}}(\z_{\rm act}, \z_{\rm pas})$ do not have the properties one would hope for.  The spin is translation-, but not supertranslation-invariant, in $\z_{\rm pas}$, and there is an ambiguity in the center of mass\footnote{For any cuts $\z_{\rm act}$, $\z_{\rm pas}$, there is a translation $\tau=\tau (\z_{\rm act},\z_{\rm pas})$, unique up to multiples of the energy--momentum, for which the mass-moments $\bm{M_{ab}}(\z_{\rm act},\z_{\rm pas}+\tau )$ vanish.  If we interpret the vanishing of the mass-moments as giving the center of mass, we find then that for each active cut there is an infinite-dimensional family of centers of mass, one $\z_{\rm pas}+\tau$ (modulo translations along $\bm{P^a}$) for each $\z_{\rm pas}$.  These different centers of mass are all mutually relatively supertranslated.}\cite{ADH2021}.)
Related to this, the infinite-dimensional dependence on $\z_{\rm pas}$ is in tension with the the hope that the angular momentum at a fixed $\z_{\rm act}$ should be a few key kinematic quantities.

Because of these questions, 
when one tries to apply the BMS-charge definitions, for instance, to compute the angular momentum emitted in an interval of radiation, one searches for low-dimensional preferred families of cuts $\z_{\rm act}$, $\z_{\rm pas}$ naturally determined by the circumstances, to try to have as firm an interpretational basis as possible.

I should remark that Dray and Streubel did make use of Penrose's quasilocal twistor ideas in defining their charges, and indeed there are overlaps between their results and the twistor ones.  For Minkowskian regimes (which do have preferred cuts), they agree, but in more general circumstances they differ.

\subsubsection{Newman-type approaches}

Approaches like Newman's involve selecting a family of preferred cuts, or closely related structures, with appealing geometric properties.  
In effect, this family serves as (or encodes) the choice of possible origins for angular momentum.
In some of these approaches the family is to be four-dimensional; in some, it is to be one-dimensional (and represent the system's center-of-mass world-line).  In all cases, it is defined in highly nonlinear terms, and depends on the radiation.

These points mean that the space of origins is not universal.  It may not be possible to even define angular momentum except at a restricted set of cuts (or related structures), and it is not {\em a priori} clear how to compare different such angular momenta.  (To my knowledge, this issue has so far only been addressed at the infinitesimal level.)  On the other hand, the structures involved do code important parts of the dynamic space--time geometry, in deep ways.

Although the twistor approach I will describe will differ essentially in that it will be universal, it will connect with Newman's ideas in an important way.  Newman and Winicour pointed out that, in special relativity, one could interpret spin as a displacement of the center of mass into the complex \cite{NW1974b}; general-relativistically Newman considered complexifying $\scrif$ and its cuts in his theory of ${\mathcal H}$-space \cite{Newman1976}.  We will be led to a view related to these.

\subsection{The twistor approach}

I have emphasized the origin-dependence of angular momentum as the source of the difficulties in extending it to general relativity; in conventional formulations of special relativity, the angular momentum is a function $M_{ab}(x)$ of the point $x$ in Minkowski space, but there is no invariant four-dimensional affine space of origins in a general Bondi--Sachs space--time.

However, the specific way the problem of origins has appeared in this discussion --- a failure to find what would be an asymptotic Minkowksi space --- is a consequence {\em not just} of the passage from special to
general relativity, but also of 
{\em the particular mathematical formulation of special-relativistic physics which has been taken as the template for the gravitational case} --- the calculus of tensor fields, with space--time the basic object.
If, instead of that, we use Penrose's twistor theory (which at the special-relativistic level is 
equivalent to conventional theory, but whose basic object is the space $\T$ of spinors of the conformal group), the passage to the gravitational case is natural and unproblematic.\footnote{The situation is very much an example of the differing insights offered by competing mathematical and physical models that Feynman
discussed in his Messenger lectures \cite{Feynman1967} (pp. 168 ff.).}

Special-relativistic twistor space $\T$ is a four-complex-dimensional vector space, equipped with certain structures (a pseudo-Hermitian form, an alternating form and an ``infinity twistor'').  From these, it is possible to recover Minkowski space with its metric, and so special-relativistic physics can be recast in twistor terms \cite{PR1986}.

The special-relativistic twistors themselves (the elements of $\T$) can be interpreted in different ways, in particular as certain spinor fields but also as geometric structures on Minkowski space.  What will be most important here is that there is a distinguished class of twistors, the {\em real} twistors (those with vanishing norm), which can be interpreted as pairs $Z=(\gamma ,\pi_{A'})$ of null geodesics $\gamma$ and tangent spinors $\pi_{A'}$.\footnote{A spinor $\pi_{A'}$ is tangent to $\gamma$ if the associated null vector ${\overline\pi}^A\pi^{A'}$ is tangent to $\gamma$.}

This structure turns out to arise naturally when we formulate angular momentum spinorially.  Recall that each vector index $a$ becomes a pair $AA'$ of spinor indices, and that the metric has the spinor form $g_{AA'BB'}=\epsilon_{AB}\epsilon_{A'B'}$ for a distinguished skew form $\epsilon_{AB}$.  (Primed indices signify conjugate spinors.)
The skewness of the angular momentum implies
\begin{eqnarray}
M_{AA'BB'} = \mu_{AB}\epsilon_{A'B'}+\mu_{A'B'}\epsilon_{AB}\, ,
\end{eqnarray}
where $\mu_{A'B'}=\mu_{B'A'}$ is the {\em angular momentum spinor}.  (The choice of $\mu_{A'B'}$ rather than $\mu_{AB}$ is conventional, as is the choice of a primed tangent spinor $\pi_{A'}$ rather than an unprimed one.)   Then:
\begin{quote}
{\em For a null geodesic $\gamma$ with tangent spinor $\pi _{A'}$, the component $\mu ^{A'B'}\pi _{A'}\pi _{B'}$ of the angular momentum is constant along $\gamma$.}
\end{quote}
This suggests that {\em we take the real twistors as origins,} rather than the points in space--time.  
More precisely, each real spinor carries both origin information (the null geodesic $\gamma$), and information about the choice of component (the spinor $\pi_{A'}$).
The angular momentum is naturally a function $A(Z)$ on the real twistors.

How can we carry this over to general relativity?
Clearly, the 
concept of pairs $Z=(\gamma ,\pi_{A'})$ applies to the Bondi--Sachs setting (requiring the null geodesics $\gamma$ to meet $\scrif$), so we have at least a natural space of real twistors in their geometric interpretation.  
On the other hand, Penrose's quasilocal twistor construction, applied at any cut $\z$ of $\scrif$, produces a twistor space $\T(\z)$, whose elements are defined as certain spinor fields on $\z$, and also a formula for $A_\z$ as a function on $\T (\z)$ \cite{Penrose1982,PR1986}.
To bring these strands together, one shows that all the twistor spaces $\T (\z)$ are canonically identifiable to a space $\TT$, and that the real twistors embed naturally in $\TT$ \cite{ADH2007}.\footnote{Although we will not need a detailed analysis of the space $\TT$ here, some comments may help avoid potential confusion.  Penrose's quasilocal twistors do not generically ``integrate up,'' as the cut $\z$ is varied, to spinor fields on $\scrif$, and so the elements of $\TT$ are not spinor fields, in the ordinary sense, on $\scrif$.  (Rather, each element of $\TT$ defines data which would determine an element of $\T(\z )$ on any cut $\z$.)   
The space $\TT$ is not a complex vector space, but a manifold with certain weak singularities.  A choice of cut $\z$ determines a complex-vector-space structure on $\TT$, this structure being Penrose's $\T (\z)$, but the structure does depend on the cut.}

The result 
is a formula $A_\z(Z)$ for the angular momentum at a cut $\z$ about the twistor $Z=(\gamma ,\pi_{A'})$, which codes both the origin and the choice of component.  
(The twistor $Z$ need not be specially related to the cut $\z$.)

Some consequences of this are worth pointing out:

(a) In passing from events to null geodesics, the concept of an origin for the measurement of angular momentum is delocalized in space--time, but it becomes localized in twistor space.

(b) The function $A_\z(Z)$ combines the energy--momentum and angular momentum syncretically, for as the twistor $Z$ is varied both of these are determined.  In the special-relativistic case, this is a compact reflection of the physical principle that these quantities together form a Poincar\'e-covariant moment map (in the language of symplectic mechanics).  
In passing to general relativity, the twistor treatment keeps a close connection between the angular momentum and energy--momentum.

(c) Since the space of pairs $(\gamma ,\pi _{A'})$ is evidently BMS-invariant, it is universal. Thus there is no difficulty at all in comparing angular momenta at different cuts --- 
they are functions on the same space.  In this sense, there is no supertranslation problem.

(d) On the other hand, because the space of real twistors has a weaker structure in the general-relativistic case than in the special-relativistic one, we do need to elucidate the interpretation of the twistorial angular momentum.  In doing this, we will see that supertranslation issues --- for example, comparing two, relatively supertranslated, Minkowskian regimes --- enter differently than in BMS-based approaches, and also that the spin and center of mass acquire qualitatively new general-relativistic corrections.

(e)  Certain additional structure is provided by the twistor construction (analogs, for $\TT$, of the reality structure, alternating form and infinity twistor on $\T$), and these are used in working out the detailed properties of the angular momentum.  

\subsection{Interpretation and results}

I have explained how the angular momentum can natural be represented as a function, not on space--time points and component indices, but on real twistors $(\gamma ,\pi_{A'})$.  In special relativity, those embed in a twistor space $\T$ with a natural linear structure, and that allows a recovery of the usual Minkowskian treatment.  In general relativity, however, even the manifold of real twistors has a weaker structure, because of the supertranslational freedom in shifting generators of $\scrif$ relative to one another.  This means that in general no asymptotic ``Minkowski space of origins'' exists,  and we have to develop an interpretation of the twistorial angular momentum which does not rely on this concept.

\subsubsection{Minkowskian regimes}

The simplest case is a Minkowskian regime.  There, the twistor structures actually {\em are} equivalent to the Minkowski ones, with the points of Minkowski space $\MM$ being the good cuts.  Also $A_\z(Z)$ is entirely independent of the choice of cut in the regime (even if $\z$ is not a good cut), so the energy--momentum and angular momentum those of a special-relativistic system on $\MM$.  (These points are consequences of the discussion in refs. \cite{Penrose1982,PR1986}.)

Next, consider the case of two, relatively supertranslated Minkowskian regimes, say $\RR$, $\acute\RR$.  
There is no difficulty at all in simply {\em comparing} the angular momenta $A_\RR$ and $A_{\acute\RR}$, for they are both functions on the same twistor space $\TT$.  
However, the linear structures $\T (\RR )$ and $\T(\acute\RR)$ will be different, so, while the angular momentum for each regime will appear as a quadratic form with respect to that regime's preferred linear structure, it will appear as a more complicated function when referred to the other regime's linear structure.  There will be no simple relation between the Minkowski spaces $\MM(\RR )$ and $\MM(\acute\RR )$.  The angular momentum $A_\z(\RR )$ will not appear as a special-relativistic angular momentum on $\MM (\acute\RR )$, and vice versa.
The need, then, is to give a physically comprehensible interpretation of $A_\RR-A_{\acute\RR}$, or equivalently to express $A_{\acute\RR}$ in terms of the Minkowski structure $\MM(\RR )$ and vice versa.

We will find that this comparison can be made with something very close to the Minkowskian change-of-origin formula.  Twistor theory in effect interprets the supertranslation between the regimes as a {\em direction-dependent translation} (the direction in question being the same as that determined by the spinor $\pi_{A'}$ indexing the component).  Apart from this, the form of the transformation of angular momenta from $\MM(\acute\RR )$ to $\MM (\RR )$ is the same as the Minkowskian one.  

These results are attractive, and adequate for many questions of physical interest.  They differ from BMS-based ones:  the twistor transformation formula involves only energy--momentum (not supermomenta); it is an algebraic (not an integral) relation; in general, there are ``longitudinal'' contributions (terms dependent on the energy--momenta of the regimes $\RR$, $\acute\RR$, not just the difference in their energy--momenta).

But we are also interested in understanding the angular momentum in dynamic regimes. For these, we cannot rely on a Minkowskian background; we must see what other interpretations survive general-relativistically.

\subsubsection{Spin and center of mass}

When we interpret the angular momentum of a specific special-relativistic system, we almost invariably pass to its center-of-momentum frame; then the mass moments $K^a$ are the time--space components of the angular momentum, and the spin is given by the space--space components $J^a$.

It is possible to develop a parallel analysis for the twistorial angular momentum (at any cut).  That this can be done is technically remarkable, and comes from the algebraic properties of an intricate set of constraints in twistor space.  The result can be viewed as a general-relativistic counterpart of the 
formula for
\begin{eqnarray}\label{crep}
  i(J^{AA'}+iK^{AA'})_\z {\overline\pi}_A\pi_{A'}
\end{eqnarray}
as a function of the choice of asymptotically constant spinor $\pi_{A'}$.  In Minkowski space, the formula (\ref{crep}) would be the usual complex representation of the angular momentum (times $i$), as an element of the complex $j=1$ representation of the Lorentz group.  In the general-relativistic setting, we find a complex $j=1$ term, but we also find $j\geq 2$ terms, which are $M\lambda$, where $M$ is the mass and $\lambda$ is the angular potential for the shear ($\sigma=\eth^2\lambda$).\footnote{Because many of the quantities of interest are spin-weighted functions, I denote the multipole by $j$ rather than $\ell$.}  

The interpretations of the $j\geq 2$ terms turn out to be satisfying.  For the center of mass, the $\Re\lambda$ term gives a supertranslation which has the effect of canceling any gauge contributions in the choice of active cut.  (For instance, if $\Im\lambda =0$, the effect of the $\Re\lambda$ term is to make any shear at $\z$ appear as if it is due to $\z$ being a bad cut in a Minkowskian regime, and to define the center of mass in a natural way using the good cuts.)
The result is a well-defined center of mass which appears to directly reflect the physics of the situation.  (Compare Ref. \cite{ADH2021}.)
And if one interprets $iM\Im\lambda$ as a complex supertranslation,
one finds a direct general-relativistic version of a special-relativistic result of Newman and Winicour \cite{NW1974b}, that spin can be interpreted as a displacement of the center of mass into the complex.  

We find then, that there are good physical reasons to accept the shear, or its angular potential $\lambda$, as $j\geq 2$ contributions to the general-relativistic angular momentum.  This means that general relativistic angular momentum should be regarded as unifying the special-relativistic ($j=1$) contributions with the shear.

\subsubsection{Strength of the effects}

The Bondi--Sachs formulas tell us that the rate of energy--momentum loss is proportional to $|\dot\sigma|^2$ --- it is purely quadratic in the gravitational radiation.  This provides an important limitation on radiative effects.

The situation for angular momentum is more complicated.  The reason is that, general-relativ\-ist\-ically both power and $\dot\sigma$ are dimensionless, whereas torque has dimension mass (or length).  This means that something must set the scale for the rate of angular momentum loss, and there are two quantities at $\scrif$ which might be naturally expected to do so:  the curvature coefficient $\psi_2$, and the shear $\sigma$. 
Depending on these quantities' relative sizes, and how each combines with the radiation $\dot\sigma$, various behaviors are possible.  The matter is further complicated by the gauge character of $\sigma$.

Let us consider the change in angular momentum between two Minkowskian regimes, and suppose (at first) no matter is present near $\scrif$.  If there is {\em no} supertranslation offset between the regimes, we find the change in angular momentum can be expressed purely as a quadratic form in the radiation field, and indeed agrees with the formula given by BMS-based approaches.  In particular, there is no ``longitudinal'' contribution.

If there {\em is} a supertranslation mismatch, we find additional terms (not present in the BMS-based approaches):  one first-order, and one third-order in the radiation field.  Each of these terms is explicitly proportional to the mismatch.  The first-order one is also proportional to $\psi_2$ (a ``longitudinal'' contribution), whereas the third-order one is involves only the radiation and is proportional to what might be called the emitted energy aspect $\int _{u_0}^{u_1}|\dot\sigma|^2\, du$.

Because of the different analytic forms of these terms, and the relative freedom in $\psi_2$ and $\sigma$, there is no obvious concise broad statement one can make about which dominates, even in the case of a uniformly-weak radiation limit.  (For instance, although formally the first-order term might be expected to dominate in those cases, we have seen that it vanishes if there is no supertranslation mismatch.)  But there are some important observations we can make.

The positive-energy theorems bound what I have called the emitted energy aspect, and thus limit the contribution of the third-order term (to of the order $M$ times a quantity involving the pointwise suprema of 
$|\Delta\lambda|$, $|\eth\Delta\lambda|$).

The case of a supertranslation mismatch is the generic one.  If we imagine fixing a non-zero $\Delta\lambda$ , then in a formal sense for sufficiently weak radiation fields (small $\dot\sigma$) the first-order term will dominate, and this will be true in many practical cases.  In particular, it will describe straightforward models of the asymptotic linearized gravitational fields of special-relativistic systems.  

Consider a special-relativistic system of localized bodies.  In the approximation that the interactions are also localized, so that the bodies may scatter, fission and combine, but in between they are freely falling and not interacting, the first-order gravitational field computed from them will give Minkowskian regimes on $\scrif$ when there is no interaction, but changes in $\sigma$ from one of these regimes to another.  Twistor theory gives changes in angular momentum.
In the quadrupole approximation, we find for this first-order effect
\begin{eqnarray}
\Delta \bm{J^a}\Bigr|_{\text{first-order, quadrupole}}
 &=&\frac{4}{15G} \bm{\epsilon^{ab}{}_c\psi^{cd}}\Delta\bm{\lambda_{db}}
\end{eqnarray}
in standard three-tensor notation,
where $\bm{\psi_{ab}l^al^b}$ is the quadrupole contribution to $\psi_2$ and $\Delta\bm{\lambda_{ab}l^al^b}$ is the change in the quadrupole part of $\lambda$
(with $\bm{l^a}$ the normalized null vector coding the direction).
Beside the derivation from twistor theory, a direct physical argument will be given for such terms.
So an ordinary system of particles, interacting by contact forces, emits a tiny, but first-order, amount of angular momentum in gravitational radiation with each scattering.

 \subsubsection{Two technical results}
 
There are two further results, not in themselves twistorial, but of interest for Bondi--Sachs space--times generally, and useful here. 
 
The first is a clarification of a point about Minkowskian regimes on $\scrif$; recall that I have defined these as having purely electric and $u$-independent shear.  It is commonly asserted that one can find good cuts in such regimes.  I show here that this need not be the case; one also needs the Minkowskian regime to persist for a sufficiently long interval.  (But it will also be shown that this distinction is irrelevant to the twistor approach, and the same observation will help with other approaches.)
 
The second result is the development of an explicit abstract-index treatment of asymptotically constant spinors and tensors.  
The issue here is that for many purposes one wants to focus simply on their multilinear structure, but the usual formalism brings in much further detail which may not be needed and can be distracting.  For example, an asymptotically constant spinor is, strictly speaking, defined as a certain equivalence class of spin-weighted fields.  For some purposes, we do want that detail, but often we would like to think of it as a unit (that is, an object whose detailed composition is irrelevant) $\bm{\pi_{A'}}$.

Other authors have had related concerns.  Often a basis is introduced, and then computations are done in terms of the components.  This makes the multilinearity clear, but it breaks the invariance.  The approach here is a development of that in ref. \cite{PR1986}.  I will use boldface symbols (e.g. $\bm{P_a}$, $\bm{\pi_{A'}}$), for the asymptotically constant spinors and tensors as units, whereas ordinary tensor or spinor fields on space--time will be lightface.

\subsection{Plan of the paper}

The next section reviews the spinorial treatment of angular momentum in special relativity, and section III give the results we will need from special-relativistic twistor theory.
Section IV begins the passage to general relativity, giving the definition of asymptotically constant spinors and their tensor algebra.
Section V recapitulates the key definitions and formulas from twistor theory at $\scrif$.

Section VI treats the angular momentum of Minkowskian regimes, including the comparison of relatively supertranslated ones.  The distinction between Minkowskian and strongly Minkowskian regimes is established there, too.  Section VII points out that, although an overall Minkowski space of origins does not exist in general circumstances, for each fixed asymptotic direction, corresponding to a choice of component of the angular momentum, 
there exists a well-defined model of Minkowski space modulo translations in that direction.

Section VIII derives the results for spin and center of mass.  

Sections IX, X, and XI discuss the emission of angular momentum in terms of the order of the gravitational radiation field involved, the first one giving a preliminary discussion, and the next two giving the first- and third-order effects, which are the ones not found in the BMS-based treatments.

Section XII is given to discussion.  An appendix outlines the connection between the asymptotic structure used here and Bondi--Sachs space--times. 

{\em Notation, conventions and background.}
This paper assumes a familiarity Penrose's null infinity and the Geroch--Held--Penrose version of the spin-coefficient formalism \cite{PR1986}.  It does review (in section II) the essential algebraic properties of two-component spinors, and the appendix outlines the properties of future null infinity in connection with Bondi--Sachs space--times.
%
I have not assumed any knowledge of twistor theory.  The twistor {\em ideas} essential to this paper are explained, but for the most part known formulas are taken from the literature.

These matters, and all necessary background material, not otherwise noted, will be found in Penrose and Rindler \cite{PR1984,PR1986}, whose notation and conventions are followed. 
The metric signature is $+{}-{}-{}-$ and $\epsilon_{txyz}=+1$ in a right-handed orthochronous frame.  The speed of light is unity, but Newton's constant $G$ is written explicitly.  

The symbol $\oint$ denotes an integral over a cut of $\scrif$ with respect to the area form of the Bondi system.

In many places, we will be concerned with deriving expressions for symmetric bilinear forms, most often the angular momentum twistor $A_\z(Z,\acute Z)$.  Such a form can always be recovered from the associated quadratic form $A_\z(Z)=A_\z(Z,Z)$ via the polarization identity $A_\z(Z,\acute Z ) =(1/4)((A_\z(Z+\acute Z,Z+\acute Z) -A_\z(Z-\acute Z ,Z-\acute Z))$, and this will be used without comment.

\section{Special relativity, spinors and angular momentum}

The problems with treating angular momentum are associated with its position-dependence.  Ultimately, this issue will be resolved by using the following special-relativistic property:  {\em if} one changes position along a given null geodesic, {\em and} examines the components selected by spinors compatible with the null tangent to the geodesic, the result is invariant.  
I will show this algebraically here, reviewing along the way some basics of two-component spinors.  This section is entirely special-relativistic.

\subsection{Preliminaries}

Recall that there are two, complex-conjugate, spin spaces $\Spin ^A$ and $\Spin ^{A'}$, each two-complex-dimensional.
Each vector index $a$ can be converted to a pair $AA'$ of spinor indices by means of the Infeld-van der Waerden symbols $\sigma _a{}^{AA'}$, $\sigma ^a{}_{AA'}$.  We thus write $v^{AA'}$ and even $v^a=v^{AA'}$ for a vector $v^a$ without comment.  
The null vectors are precisely those which can be written in spinor form as $\pm{\overline\pi}^A\pi ^{A'}$, the plus or minus sign according to whether the vector is future- or past-directed.  (The overbar indicates conjugation.  The choice of $\pi _{A'}$ as the unconjugated spinor is not important but is compatible with twistor conventions.)
If such a null vector $v^a$ is the tangent to a null geodesic, any primed spinor proportional to $\pi _{A'}$, or any unprimed spinor proportional to ${\overline\pi}_A$, is said to be tangent to the geodesic.

Each spin space is equipped with a non-degenerate skew form: $\epsilon _{AB}$ and $\epsilon _{A'B'}$.  (One should, strictly speaking, write ${\overline\epsilon}_{A'B'}$ for the latter, but for certain common spinor quantities, when there is no danger of confusion, it is customary to omit the overbar.)
We denote by $\epsilon ^{AB}$ minus the inverse of $\epsilon _{AB}$; then we have
$\epsilon _{AB}\epsilon ^{AB}=2$, and spinor indices are raised an lowered with the conventions $\alpha ^A=\epsilon ^{AB}\alpha _B$, $\alpha _A=\alpha ^B\epsilon _{BA}$.  Note that the skewness of $\epsilon _{AB}$ implies $\alpha ^A\alpha_A =0$.
The metric tensor has the spinor form $g_{AA'BB'}=\epsilon _{AB}\epsilon _{A'B'}$.  

Because spin-space is only two-dimensional, there are considerable simplifications in the symmetry properties of quantities with a number of spinor indices.  If a quantity $\phi _{\cdots AB\cdots}$ is skew on two indices (both primed, or both unprimed), it is necessarily proportional to the corresponding $\epsilon _{AB}$ (or $\epsilon _{A'B'}$).  Thus any spinor may be decomposed over its indices of a given type (primed or unprimed) into a sum of totally symmetric spinors and epsilon factors.

An important application of this is to space--time two-forms.  If $F_{ab}=-F_{ab}$, then its spinor form $F_{AA'BB'}$ must be a sum or parts symmetric on $A$ and $B$ but skew on $A'$ and $B'$, and a term skew on $A$ and $B$ but symmetric on $A'$ and $B'$.  Thus we have
\begin{eqnarray}
  F_{AA'BB'}=\phi _{AB}\epsilon _{A'B'} +\phi _{A'B'}\epsilon _{AB}\, ,
\end{eqnarray}
where $\phi _{AB}=\phi _{BA}$, $\phi _{A'B'}=\phi _{B'A'}$ (and these are Hermitian conjugates if $F_{ab}$ is real).  
The tie between spinors and orientation is conveniently coded in this decomposition.
The two complex quantities $F^{+}_{AA'BB'} = \phi _{A'B'}\epsilon _{AB}$, $F^{-}_{AA'BB'}=\phi _{AB}\epsilon _{A'B'}$ are, respectively, self- and anti-self-dual:
\begin{eqnarray}
  {}^*\!F^{\pm}_{ab}&:=&(1/2)\epsilon_{ab}{}^{cd}F^{\pm}_{cd} = \pm iF^{\pm}_{ab}\, .
\end{eqnarray}

Finally, the conventions for components.  Often one works with a dyad $\omicron^A$, $\iota^A$ normalized to $\omicron_A\iota^A =1$.  The components of a spinor $\xi^A$ with respect to this are set by $\xi^A=\xi^0\omicron^A+\xi^1\iota^A$.  It is important to appreciate that this convention holds, whether the spinor would naturally be defined with its index up or down, for one finds $\xi_0 = -\xi^1$ and $\xi _1=\xi^0$.  (To see this, note that the normalization condition gives $\epsilon_{01}=1$, whence $\epsilon^{01}=1$.)

\subsection{Angular momentum spinorially}

We will be especially interested in generalizing the special-relativistic angular momentum, which is a position-dependent two-form:
\begin{eqnarray}
  M_{ab}(x + \tau) =M_{ab}(x) +P_a\tau_b-P_b\tau_a\, ,
\end{eqnarray}
where $P_a$ is the energy--momentum.  In spinor terms, one conventionally writes
\begin{eqnarray}
  M_{ABA'B'} =\mu_{AB}\epsilon _{A'B'}+\mu _{A'B'}\epsilon _{AB}\, ,
\end{eqnarray}
where $\mu _{AB}$ is symmetric (and position-depedent).  The three complex components of this spinor code the six relativistic angular momentum components, three of these being ordinary angular momentum and the other three the mass-moments, which determine the center of mass.  

The key algebraic property of the angular momentum will be this:
Let us consider a change of position which is purely null:  $\tau^{AA'}= {\overline\pi}^A\pi ^{A'}$.  Suppose we {\em also} look at components of the angular momentum selected by the form $\pi ^{A'}\pi^{B'}\epsilon ^{AB}$.  Then we find
\begin{eqnarray}
&&M_{AA'BB'}(x+\alpha \tau)\pi ^{A'}\pi ^{B'}\epsilon ^{AB} \nonumber\\
   &&\mspace{30mu} = M_{AA'BB'}(x)\pi ^{A'}\pi ^{B'}\epsilon ^{AB}
   \nonumber\\
   &&\mspace{120mu}
  +2\alpha P_{AA'}{\overline\pi}_B\pi_{B'}\pi ^{A'}\pi ^{B'}\epsilon ^{AB}
  \nonumber\\
 &&\mspace{30mu}= M_{AA'BB'}(x)\pi ^{A'}\pi ^{B'}\epsilon ^{AB}\nonumber\\
 &&\mspace{30mu} =2\mu _{A'B'}\pi ^{A'}\pi ^{B'}\, ,
\end{eqnarray}
in consequence of the relation
$\pi _{B'}\pi ^{B'} =0$.  Thus {\em if one considers the component $\mu _{A'B'}\pi ^{A'}\pi ^{B'}$ of the angular momentum, this is unaltered by changes of position along a null geodesic with tangent ${\overline\pi}^A\pi ^{A'}$. } 
It is easily checked that knowledge of this quantity, for all the different events $x$ and spinors $\pi _{A'}$ determines that angular momentum, and indeed the energy--momentum.

So the angular momentum can be viewed as a well-defined function on pairs $(\gamma ,\pi _{A'})$, where $\gamma$ is a null geodesic and $\pi _{A'}$ is a tangent spinor, and the energy--momentum can be recovered from this.

Relative to a  
unit future-directed timelike vector 
$t^a$, the orthogonal vectors $J^a$ and $K^a$ representing the spatial angular momentum and mass moment, about $x$, are determined by
\begin{eqnarray}
  M_{ab} (x)&=&   \epsilon_{cdab} t^c J^d     +     K_a t_b - t_a K_b\, .
\end{eqnarray}  
If we specialize to a center-of-momentum frame, so $P_a=Mt_a$, then
a short calculation shows
\begin{eqnarray}
    2it_{AB'}\mu_{A'}{}^{B'} &=&J_{AA'}+iK_{AA'} 
  \, ,
\end{eqnarray}
and $J^a$ will be the spin, with $K^a/M$ (modulo $P^a$) the center of mass.
For this reason, the quantity $2i\mu_{A'B'}$ is often convenient to work with.

It will be helpful, for the general-relativistic case, to relate this to the view of tensors and spinors as elements of certain Lorentz representations of functions on the sphere.  
Allowing the null vector $l^a$ to represent a point on the sphere, the function
$(J_a+iK_a)l^a$ gives the familiar complex $j=1$ represention of the angular momentum.

\section{Twistors in special relativity}

For our present purposes, a special-relativistic twistor can be regarded as a spinor field $\omega^A(x)$ satisfying the {\em twistor equation:}
\begin{eqnarray}\label{tweq}
\nabla ^{A'(A}\omega^{B)}=0\, .
\end{eqnarray}
The solutions to this have the form
\begin{eqnarray}
  \omega^A(x)&=&\omega^A_0-ix^{AA'}\pi_{A'}^0
\end{eqnarray}  
for constant spinors $(\omega^A_0,\pi_{A'}^0)$, which may be regarded as the coordinates of the twistor.  
We write $Z^\alpha$ for the twistor as a whole and $\T$ for the space of twistors.  Notice that $\pi_{A'}=\pi_{A'}^0$ is  origin-independent, but $\omega^A_0=\omega^A(0)$ depends on the choice of origin.  We may call $\pi _{A'}$ the {\em direction} spinor of the twistor.  Twistor space $\T\simeq\C^4$ is the space of twistors.  

It turns out that $\T$ is the space of spinors of the conformal group of Minkowksi space, that group being 4--1 covered by $SU(2,2)$.  The pseudo-Hermitian form is $\Phi (Z) =\omega^A{\overline\pi}_A+\conj$.  (In particular, this is independent of the choice of origin.)
A twistor is said to be {\em null} or {\em real} if $\Phi (Z)=0$.  (The scale of the alternating symbol $\epsilon_{\alpha\beta\gamma\delta}$ is determined by the components $\epsilon_{AB}{}^{C'D'}=\epsilon_{AB}\epsilon^{C'D'}$.)
The conformal invariance is broken by the {\em infinity twistor} $I_{\alpha\beta}=-I_{\beta\alpha}$, given by 
$I_{\alpha\beta}Z^\alpha {\acute Z}^\beta =\epsilon^{A'B'}\pi_{A'}{\acute\pi}_{B'}$.  The subgroup of $SU(2,2)$ preserving this covers the Poincar\'e group.

A twistor $Z^\alpha$ is said to {\em pass through} a point $x$ if $\omega^A(x)=0$; such a twistor is necessarily real.  Contrariwise, if $(\omega ^A_0,\pi_{A'})$ are the coordinates of a real spinor, then (a little algebra shows) the points on it are exactly those of the form
\begin{eqnarray}\label{ngeod}
  \gamma ^{AA'}(s) &=&B^{AA'} +s{\overline\pi}^A\pi^{A'}\, ,
\end{eqnarray}
where the null impact vector
\begin{eqnarray}\label{imp}
  B^{AA'} &=& i\frac{\omega^A_0 {\overline\omega}^{A'}_0}{  
      {\overline\omega}^{C'}_0 \pi_{C'}}
\end{eqnarray}
is real in consequence of the reality of the twistor.  These equations show that the points on the real twistor are a null geodesic with tangent ${\overline\pi}^A\pi^{A'}$.\footnote{If the denominator on the right of eq. (\ref{imp}) vanishes, the null geodesic is a generator of $\scrif$.}

Any point $x$ in Minkowski space, then, has a two-complex-dimensional space of twistors passing through it, those with coordinates of the form $(ix^{AA'}\pi_{A'},\pi_{A'})$.  In fact, the two-dimensional totally real subspaces of $\T$ are in one-to-one correspondence with the points of conformally compactified Minkowski space $\MM^\#$, and the ones corresponding to Minkowski space $\MM$ itself are those whose non-trivial elements satisfy $I_{\alpha\beta}Z^\beta\not=0$, that is, $\pi _{A'}\not=0$.  
The metric can also be expressed in twistor terms.
Thus Minkowski space $\MM$ can be recovered from twistor structures.  (For the metric, see Ref. \cite{PR1986}.)

As explained earlier, in some cases we have a candidate space $\T\simeq\C ^4$ but not a pseudo-Hermitian form $\Phi$.  Then we may identify the complex two-dimensional subspaces of $\T$ with the points of complexified conformally compactified Minkowski space $\C\MM^\#$, and those whose non-trivial elements satisfy $I_{\alpha\beta}Z^\beta\not=0$ with complexified Minkowksi space $\C\MM$, but we cannot fix a real slice, and the structure is too weak to define a special-relativistic angular momentum.

The core of the analysis is that the energy--momentum and angular momentum of a special-relativistic system can be encoded twistorially.  The {\em kinematic} or {\em angular momentum twistor} has coordinate form
\begin{eqnarray}
  A_{\alpha\beta} =\left[\begin{matrix} 0& P_A{}^{B'} \\ P_B{}^{A'} & 2i\mu^{A'B'}
  \end{matrix}
  \right]\, ,
\end{eqnarray}  
where $\mu^{A'B'}$ is the angular momentum spinor with respect to the origin.  
The fact that $A_{\alpha\beta}$ {\em is} a twistor --- that it behaves properly under changes of origin --- is due to the transformation law for angular momentum.
For a real twistor $Z^\alpha$ we have
\begin{eqnarray}\label{tang}
  A_{\alpha\beta}Z^\alpha Z^\beta = 2i\mu^{A'B'}(x)\pi _{A'}\pi_{B'}\, ,
\end{eqnarray}
where $x$ is any point $Z^\alpha$ passes through.  
We saw earlier, as a sort of trick of spinor algebra, that this was independent of the choice of point along the null geodesic; now we see that this is a consequence of natural invariances encoded in the twistor formalism.  That (\ref{tang}) is a scalar --- in particular, that it is origin-independent --- comes from the compensating change-of-origin formulas for the twistor $Z^\alpha$ and the angular momentum.

More generally, if $Z^\alpha$, ${\acute Z}^\alpha$ are real twistors, then
\begin{eqnarray}
A_{\alpha\beta}Z^\alpha {\acute Z}^\beta &=& 2i\mu^{A'B'}(x_{\rm av})\pi_{A'}{\acute\pi}_{B'}
  -(i/2)P_ax_{\rm diff}^a \pi_{B'} {\acute\pi}^{B'}\, ,\nonumber\\
  &&
\end{eqnarray}
where $x^a_{\rm av}=(x^a+{\acute x}^a)/2$, $x^a_{\rm diff} =x^a-{\acute x}^a$, with $x^a$, ${\acute x}^a$, respectively, any two points in the twistors $Z^\alpha$, ${\acute Z}^\alpha$ \cite{ADH2007}.

It is also possible to recover the spin and center of mass in twistor terms.  This is critical for the interpretation of angular momentum in general relativity:  it will be the those results which will show us Bondi shear as a kind of angular momentum.

Because of this key interpretational role, I will give here the special-relativistic formula.  While this underlies essential results below, the general reader need not sort through the details here.  The points to notice are that it involves an intricate but natural set of constraints on the twistors $Z^\alpha$, ${\acute Z}^\alpha$ which appear, but that {\em these constraints do not involve any choice of space--time point}.
(In general relativity, it will be the counterparts of these constraints which bring the shear into the spin and center of mass.)

Let the twistors $Z^\alpha$, ${\acute Z}^\alpha$ be real, and also satisfy $A_{\alpha\beta}Z^\alpha Z^\beta =A_{\alpha\beta}{\acute Z}^\alpha{\acute Z}^\beta=0$.  
Let $P_a$ be the energy--momentum and $M$ the mass.
Set
\begin{eqnarray}
 \alpha &=& iM\omega^A{\overline\pi}_A/(P^{AA'}{\overline\pi}_A\pi_{A'})\, ,\\
\acute\alpha &=& iM{\acute\omega}^A{\acute{\overline\pi}}_A/(P^{AA'}{\acute{\overline\pi}}_A{\acute\pi}_{A'})\, .
\end{eqnarray}
(These symbols should not be confused with twistor indices.)  Note that the reality of the twistors implies $\alpha$, $\acute\alpha$ are real; also note that they are origin-dependent
and have dimension length.  In fact, they are the values of the Bondi retarded time at which the geodesics strike $\scrif$, in the center-of-momentum frame.  It turns out that one may satisfy the constraints, with no additional freedom, by choosing $\pi_{A'}$, ${\acute\pi}_{A'}$ and the real values $\alpha$, $\acute\alpha$ arbitrarily.

A direct computation shows
\begin{widetext} 
\begin{eqnarray}
&&\left( A_{\alpha\beta}Z^\alpha{\acute Z}^\beta /I_{\alpha\beta}Z^\alpha {\acute Z}^\beta
  \right) \left( P^{BB'} {\overline\pi}_B\pi_{B'}  \right)
  \left( P^{BB'}  {\acute{\overline\pi}}_B{\acute\pi}_{B'} \right)
 \\
 &&\quad = \frac{M}{2}\left[
  \left( J^{AA'}+i(K^{AA'}+\alpha P^{AA'})\right){\overline\pi}_A\pi_{A'} 
  \left( P^{BB'}  {\acute{\overline\pi}}_B{\acute\pi}_{B'} \right)
  - \left( J^{AA'}+i(K^{AA'}+\acute\alpha P^{AA'})\right){\acute{\overline\pi}}_A{\acute\pi}_{A'} 
  \left( P^{BB'} {\overline\pi}_B\pi_{B'}  \right)\right]
  \, ,\nonumber
\end{eqnarray}
\end{widetext}
and knowledge of this quantity, as a function of the allowable twistors $Z^\alpha$, ${\acute Z}^\alpha$, determines the angular momentum in the form $J^a+iK^a$.   
Inspection shows that the real part of this identity determines the spin $J^a$.  
The imaginary part determines the mass moment $K^a$ (or equivalently, the center of mass $K^a/M$) relative to the coordinate origin.  When we come to the general-relativistic twistor case, 
there will be no sense of a coordinate origin, but we will see that this loss of structure is compensated by a natural reinterpretation of the formula, one which provides a satisfying, and arguably compelling, definition of center of mass.

\section{Asymptotically constant spinors and tensors}\label{sec:spinsec}

When gravitational radiation is present, {\em covariantly constant} spinor and tensor fields do not exist in the limit of passage to $\scrif$:  Sachs peeling, a basic scaling behavior of the curvature, implies that there are obstructions.  There does exist, however, a concept of {\em asymptotically constant} spinor and vector fields; for instance, the Bondi--Sachs energy--momentum is an element of the space of asymptotically constant covector fields.
The definition of these fields is {\em non-local:}  it involves solving equations which are elliptic over the sphere of directions.

\subsection{Null infinity and rescalings}

For the rest of this paper, we will be working primarily at future null infinity, and the notation will be adapted accordingly.  {\em From now on, except as noted, the physical space--time and geometric quantities associated with it will be denoted with hats: 
$\widehat{M}$, ${\widehat g}_{ab}$, ${\widehat\nabla}_a$, etc.; unhatted symbols 
($M$, $g_{ab}$, $\nabla _a$, etc.) will refer to the conformal extension.}  Recall that $g_{ab}=\Omega^2{\widehat g}_{ab}$ and that $M={\widehat M}\cup\scrif$.

We will also make use of the Newman--Penrose formalism.  We recall that a normalized spinor dyad $\omicron^A$, $\iota^A$ is introduced, so that $l^{AA'}=\omicron^A\omicron^{A'}$ is the generator of affinely parameterized null geodesics ruling the $u=\const$ hypersurfaces, $m^{AA'}=\omicron^A\iota^{A'}$ is an antihilomorphic tangent to the $u=\const$ cuts, and $n^{AA'}=\iota^A\iota^{A'}$ is on $\scrif$ a null generator; the dyad is transported parallel along $l^a$.  The spin- and boost-weight covariant derivatives are $\thorn$, $\eth$, $\eth'$, $\thorn'$ in the $l^a$, $m^a$, ${\overline m}^a$, $n^a$ directions.

A few points to note:

(a) We will be working at $\scrif$ (or to first order off $\scrif$).  At $\scrif$, the direction of $\iota^A$ is BMS-invariant, but $\omicron^A$ depends on the Bondi system used.  This will underlie the invariance in the choice of certain components, below.

(b) The relative scaling of the physical and unphysical dyads will come up:  we have $\omicron^A=\Omega^{-1}{\hat\omicron}^A$, $\iota^A={\hat\iota}^A$, $\omicron_A={\hat\omicron}_A$, $\iota_A=\Omega{\hat\iota}_A$.

(c) We will work in a Bondi system for which $n^a$ is divergence-free, so the spin-coefficient $\rho'=0$ at $\scrif$.

\subsection{Asymptotically constant spinors}

We want to know what conditions should be considered to characterize an asymptotically constant spinor field $\xi^A$.  The first requirement should be that, with respect to a {\em physical} Bondi basis ${\hat\omicron}^A$, ${\hat\iota}^A$, the components of $\xi^A$ should be bounded at $\scrif$.  This means that the (rescaled) component $\xi^0$ will vanish in that limit.  That in turn means that the component $\xi^1=\xi^A\omicron_A$ will be BMS-invariant (unaffected by additions of $\iota_A$ to $\omicron_A$).  

Now let us look to which of these fields should be considered asymptotically constant.  The {\em covariant} constancy condition would be the vanishing of ${\widehat\nabla}_{AA'}\xi ^B
=\nabla _{AA'}\xi ^B - \epsilon_A{}^B \Upsilon _{XA'}\xi ^X$, but keeping all of the components of this tangent to $\scrif$ results in an overdetermined system (if gravitational radiation is present).  However, if we just look at those got from contractions with $\iota^{A'}$, we do get a two-dimensional space.  So we define
\begin{eqnarray}\label{spindef}
\Spin ^A&=&\{ \xi\mid \xi\text{ has type } \{ 1,0\},\ 
  \eth\xi =0 =\thorn'\xi \}\, ,\qquad
\end{eqnarray}  
where $\xi$ is to be thought of as the component $\xi^1$.

The definition of $\Spin^A$ just given is conventional \cite{PR1986}.  It is rather abstract, in that the spinor is represented by a single spin-weighted field.  We can however use another of the components of the covariant-constancy condition to associate to it a certain equivalence class of spinor fields:
\begin{eqnarray}\label{hatxi}
  \xi\leftrightarrow {\hat\xi}^A=\xi \iota^A -\Omega (\eth'\xi )\omicron^A\, ,
\end{eqnarray}
where ${\hat\xi}^A$ is taken to be defined on the first formal neighborhood of $\scrif$ (that is, working infinitesimally to first order off $\scrif$), and also only modulo terms proportional to $\Omega\iota^A$.
(Passing to the equivalence class is necessary for BMS covariance.)
The notation ${\hat\xi}^A$ is meant to suggest the physical field (or more properly, equivalence class of fields) corresponding to the asymptotic constant spinor.
We may define ${\hat\xi}^1=\xi$, ${\hat\xi}^0=-\eth'\xi$.
Note that ${\hat\xi}^1$ may be recovered from $\eth'{\hat\xi}^0$, as $\xi =-2[\eth ,\eth']{\hat\xi}^1 =2\eth{\hat\xi}^0$, using the commutator
\begin{eqnarray}\label{comm}
  [\eth, \eth'] \zeta = -2s\zeta
\end{eqnarray}
on spin-weight $s$ quantities $\zeta$.  

Higher-valence spinors (and so vectors and tensors) can be built up out of tensor products of the spin spaces.  If we do this literally, they become fields on products of copies of $\scrif$, satisfying certain differential relations.  
While we will sometimes take that point of view, more commonly it will be useful
to consider the restrictions of those fields to the diagonal (but enforce the differential relations derived from off-diagonal considerations), and unless otherwise noted we do this.  In particular, then, the asymptotically constant vectors are fields $v$ of type $\{ 1,1\}$ satisfying $\eth^2v =0$, $\thorn'v=0$.

Since asymptotic spinors are certain fields on $\scrif$, dual spinors would be strictly defined in terms of distributions.  However, in practice the skew form will be used to identify duals.

\subsection{Abstract notation and duals}

Most often in this paper, we are interested in asymptotically constant spinors as elements of the two-complex-dimensional vector space $\Spin^A$; revisiting their definition as fields on $\scrif$ would, for this purpose, be distracting.  It will be helpful to have a notation which reflects this.  I will use boldface:
\begin{eqnarray}
\xi\leftrightarrow \bm{\xi^A}\in\Spin^A\, .
\end{eqnarray}
As a first application of this, we note that the form (\ref{hatxi}) allows us to see that the conventional expression for the alternating form gives a well-defined limit at $\scrif$:
\begin{eqnarray}
  \bm{\epsilon_{AB}\xi^A{\acute\xi}^B} = \xi\eth'\acute{\xi} -\acute{\xi}\eth'\xi\, ,
\end{eqnarray}  
and (applying $\eth'$) one sees this is constant.  This defines $\bm{\epsilon_{AB}}$.

We will consider dual spinors to be defined by the lowering operation:  $\bm{\xi_A}=\bm{\xi^B\epsilon_{BA}}$.

\subsection{Dyad from a spin-frame at $\scrif$}

It is sometimes useful to introduce a dyad $\bm{\omicron^A}$, $\bm{\iota^A}$ associated with the spin-frame $\omicron^A$, $\iota^A$ at a particular generator $\gamma$ of $\scrif$.  We let $\bm{\omicron^A}$, $\bm{\iota^A}$ be the 
elements
$\omicron^1$, $\iota^1$ in $\Spin^A$ with 
\begin{align}
    \omicron^1\Bigr|_\gamma &=0\, ,&\quad 
              \eth'\omicron^1\Bigr|_\gamma &=-1\, ,\\
    \iota^1\Bigr|_\gamma &=1\, , &\quad 
            \eth'\iota^1\Bigr|_\gamma &=0\, .
\end{align}    
These are a properly normalized basis, and $\bm{\iota^A}\omicron_B -\bm{\omicron^A} \iota_B$ provides an isomorphism from the local spin-space at a point on $\gamma$ to $\Spin^A$.  We have then $\bm{\xi^A}=\bm{\xi^0\omicron^A}+\bm{\xi^1\iota^A}$ with
$\bm{\xi^0} ={\hat\xi}^0\Bigr|_\gamma = -\eth'\xi\Bigr|_\gamma$, $\bm{\xi^1}={\hat\xi}^1\Bigr|_\gamma=\xi\Bigr|_\gamma$.

Then $\bm{t^{AA'}}=(1/2)\bm{\omicron^A\omicron^{A'}}+\bm{\iota^A\iota^{A'}}$ is the timelike vector associated with the Bondi frame.
This asymptotically constant vector is (by definition) the same as the field $(1/2)\omicron^1 \overline{\omicron^1} +\iota^1 \overline{\iota^1}$.  That field has 
the constant value unity,\footnote{This can be seen by noting the field will be determined by its
 value and its $\eth$, $\eth'$, $\eth\eth'$ derivatives at any given point, and the derivatives will vanish at the defining generator.  The field is that denoted $A$ in \cite{PR1986}, their eq. (9.6.27) ff.} reflecting the normalization of the spin-frame to the Bondi system.

The map $\Spin^A\to\Spin^{A'}$ given by ${\bm\xi^A}\mapsto \bm{t^{AA'}\xi_{A}}$
will come up.  The image is
$-(1/2)\bm{\omicron^{A'}}\bm{\xi^{1}} +\bm{\iota^{A'}}\bm{\xi^{0}}$, the field whose value at $\gamma$ is $\bm{\xi^{0}}$ and whose $\eth$ derivative is $(1/2)\bm{\xi^{1}}$ there.
Using the commutator relation (\ref{comm}), one can verify the field is
\begin{eqnarray}\label{tesk}
  -\eth'\xi\leftrightarrow \bm{t^{AA'}\xi_{A}}\, .
\end{eqnarray}

\section{Results from twistor theory}

I give here the key formulas from the twistor theory of angular momentum which we will need.  
See refs. \cite{PR1986,ADH2007} for derivations.

\subsection{Twistors at $\scrif$}

Twistors in Minkowksi space were defined as solutions to the twistor equation (\ref{tweq}).  This equation is conformally invariant, and so extends meaningfully to $\scrif$, but it is generally overdetermined there and has only the trivial solution.  The general-relativistic twistor space $\TT$ at $\scrif$ is constructed by keeping only certain, geometrically distinguished, components of that equation, restricted to certain, geometrically distinguished, submanifolds of $\scrif$.

As mentioned above, the twistor space $\TT$ is a bundle over asymptotic spin space $\Spin_{A'}$.  The base space may be identified by taking over the components
\begin{eqnarray}\label{omz}
\eth'\omega^0= 0\, ,\qquad \thorn'\omega^0 =0\, ,
\end{eqnarray}
of eq. (\ref{tweq}),
so $\omega^0$ represents an asymptotic primed spinor\footnote{The reader may be concerned that it is the $0$ component of the spinor which enters here, whereas it was the $1$ component at the beginning of section IVB.  The resolution is that the spinor field $\omega^A$ would not be asymptotically constant, even for Minkowski space--time.} which is identified as $i\bm{\pi^{A'}}$.  If $\omega^0$ is not identically zero, there will be a unique generator $\gamma =\gamma (\omega^0)=\gamma (\bm{\pi_{A'}})$ of $\scrif$ on which it vanishes.  (We will not need the case of identically zero $\omega^0$; see ref. \cite{ADH2007} for that.)

The remaining degrees of freedom for the twistor are specified by certain data on $\gamma$, essentially a complex affine two-space which can be thought of as the values one would like to assign to $\omega^1$ and $\eth'\omega^1$ on that generator.  Given any cut $\z$, one can use these data and Penrose's quasilocal twistor equation to determine a field $\omega^1$ on $\z$.\footnote{But in general this field is strongly cut-dependent, so for example, two cuts which agree in a neighborhood of $\gamma$ but disagree elsewhere will in general have different quasilocal $\omega^1$ fields in the neighborhood; only their data $\omega^1$, $\eth'\omega^1$ {\em at} $\gamma$ will agree.}  For simplicity, I will work with the $u=\const$ cuts of a given Bondi system; that will be enough for the results of this paper.  Then we have well-defined fields $\omega^1$ on $\scrif$.

Fix, then, a Bondi coordinatization, and a potential $\lambda$ for the shear relative to to that, so that $\sigma=\eth^2\lambda$.  (It is often convenient to fix the freedom in $\lambda$ so that it has vanishing $j=0$ and $j=1$ terms, but that is not necessary here.)
The field $\omega^1$ has the form
\begin{eqnarray}\label{omone}
  \omega^1 &=&\omega^0\eth\lambda -\lambda\eth \omega^0
    +\alpha\eth\omega^0 +\beta\overline{\omega^0}\, ,
\end{eqnarray}
where $\alpha$ and $\beta$ (functions only of $u$) evolve according to
\begin{eqnarray}
  \alpha(u)-\alpha (u_0) &=& u-u_0   +\lambda (u,\gamma )-\lambda (u_0,\gamma)
  \label{alphaeq}\\
  \beta(u)-\beta(u_0) &=&\left[ \frac{\eth\omega^0}{\overline{\eth\omega^0}}
    \left(\eth'\lambda (u)-\eth'\lambda(u_0)\right)\right]\Bigr|_\gamma\, .
    \label{betaeq}
\end{eqnarray}    
(These formulas are derived from certain components of the the twistor equation (\ref{tweq}).)
Each twistor is then specified by four complex degrees of freedom $(\bm{\pi_{A'}},\alpha (u_0),\beta(u_0))$, with the fiber coordinates $\alpha(u_0)$, $\beta (u_0)$ depending on the Bondi system.

The infinity twistor is given by
\begin{eqnarray}
  I(Z,\acute Z ) &=&\bm{\epsilon^{A'B'}\pi_{A'}{\acute\pi}_{B'}}\, ,
\end{eqnarray}
the skew form on the base space $\Spin_{A'}$.

A twistor is real iff $\Im (\lambda(\gamma )-\alpha)=0$.  Note that this condition is independent of $u$, and also that it is highly nonlinear if $\Im\lambda$ has $j\geq 2$ terms, because $\Im\lambda(\gamma ) = \lambda (\gamma (\bm{\pi_{A'}}))$.
Recall that the $j\geq 2$ parts of $\Im\lambda$ can be regarded as an obstruction to identifying $\z$ with a cut of a Minkoswkian $\scrif$; this is reflected in the strong nonlinearity of the reality condition for $\TT$.  This will play an essential role in the interpretation of the twistorial angular momentum.

\subsection{Twistor fields and tangents at $\scrif$}

Recall that we want to view real twistors in two ways:  as fields, and also as null geodesics together with tangent spinors.  In the previous subsection, formulas for the fields (for a given Bondi coordinatization) were given; here I go over the tangent spinors.

It will help to note that there is a certain complementarity between the concepts of angle and displacement at finite points versus at infinity.  Two null geodesics which are asymptotically abreast and parallel (approach {\em zero angle, in the sense of the physical space--time,} in the asymptotic regime) will reach the {\em same point} at $\scrif$.  On the other hand, those geodesics will, in the rescaled space--time, enter that point with {\em distinct tangents} --- they will have a {\em non-trivial angle, in the sense of the rescaled space--time}.  In terms of the data available to us at $\scrif$, then, the point at which the geodesic strikes codes the {\em physical direction}, whereas the tangent to the geodesic at that point has information about the {\em impact vector}.

The projection $\bm{\pi_{A'}}$ already introduced codes the generator $\gamma (\bm{\pi_{A'}})$ at which the twistor strikes $\scrif$, and hence the {\em physical direction.}  This is defined whether the twistor is real or not.

If a twistor is real, then the field $\omega^A$ vanishes at a unique point $p$ on $\gamma(\bm{\pi_{A'}})$.
The tangent spinor to the twistor is denoted $\pi_{A'}$; it is a local spinor at $p$, 
and should be sharply distinguished from $\bm{\pi_{A'}}$ (an element of $\Spin_{A'}$).\footnote{The near-conflict in notation is unfortunate, but grows out of standard choices in the literature.}
The components of $\pi_{A'}$ are
\begin{eqnarray}
 \pi_{0'}= i\eth'\omega^1\Bigr|_{\gamma(\bm{\pi_{A'}})}\, ,\qquad 
  \pi_{1'}= i\eth\omega^0\Bigr|_{\gamma(\bm{\pi_{A'}})}\, .
\end{eqnarray}
The ratio of these tells us which of the null geodesics through $p$ is selected.

\subsection{The angular momentum twistor}

The angular momentum twistor is given by
\begin{eqnarray}\label{kintwist}
A_u(Z) &=&\frac{-i}{4\pi G}\oint \{ \psi_1(\omega^0)^2
  +2\psi_2 \omega^0{\omega}^1
    +\psi_3(\omega^1)^2\}\nonumber\\
&=&\frac{-i}{4\pi G}\oint \{ \psi_1(\omega^0)^2 
  +2(\psi_2+\sigma\dot{\overline\sigma}) \omega^0\omega^1)\}\qquad
\end{eqnarray}    
(the equality following from an integration by parts).

The Bondi--Sachs energy--momentum is
\begin{eqnarray}
\bm{P^{AA'}{\overline\pi}_A\pi_{A'}} &=&
  \frac{-1}{4\pi G}\oint (\psi_2+\sigma\dot{\overline\sigma})\omega^0\overline{\omega^0}\, .
\end{eqnarray}
By writing $\omega^0 = i\bm{\pi^{A'}} = i\left(\bm{\pi^{0'}\omicron^{A'}}
+\bm{\pi^{1'}\iota^{A'}}\right)$, using the dyad of the previous section, one could compute the components of $\bm{P^{AA'}}$ explicitly.  The same principle will apply in parallel cases below.

\section{Angular momentum in Minkowskian regimes}

This section develops the interpretation of angular momentum in Minkowskian regimes, including multiple, relatively supertranslated, ones.  This is the commonest and one of the most important situations in which difficulties in treating angular momentum have arisen, especially the case of comparing the angular momenta of a system before and after the emission of gravitational radiation.  We can bring in much more special-relativistic structure than will be available in the general, dynamic, case.

First, a point which has lingered for some time in the analysis of Bondi--Sachs space--times is cleared up.  It is shown that a regime at $\scrif$ which is Minkowskian need not admit good cuts;  regimes which do admit good cuts will be called {\em strongly Minkowskian}.  
On the other hand, it is noted that we may as a mathematical fiction extend any Minkowskian regime to a strongly Minkowskian one.  

In any one Minkowksian regime $\RR$, twistor theory naturally defines a Minkowski space $\MM (\RR )$ and, relative to that, a special-relativistic energy-momentum $\bm{P_a}^\RR$ and angular momentum $\bm{\mu^{A'B'}}_\RR$.  In this case, the points of the Minkowski space are identifiable, as usual, with the good cuts in the regime (or its strongly Minkowskian extension).  The energy--momentum and angular momentum are strictly independent of the choice of active cut $\z_{\rm act}$ within $\RR$.  These results are direct consequences of Penrose's quasilocal construction.

But when we come to compare the angular momentum in two or more Minkowskian regimes $\RR_j$, we must go beyond familiar special-relativistic structures, for the Minkowski spaces
$\MM (\RR_j )$ will generally be relatively supertranslated.  We will show that an almost-Minkowskian transformation law holds:  
in referring the angular momentum of one regime to another, one gets a term whose form has the usual cross-product algebraic structure, but where the translation involved is direction-dependent (that is component-dependent), this direction-dependent translation being essentially the gradient of the supertranslation.

This change-of-regime term is thus a (direction-dependent) generalization of the special-relativistic one; it is quite different from the corresponding formula in BMS-based approaches (a change of passive origin), since it does not bring in the supermomenta and it is algebraic rather than integral.  It is also what has been called {\em longitudinal,} that is, it brings in more than just the difference in the initial and final energy--momenta.

Because the change-of-regime term is linear in the supertranslation, it will contribute at {\em first order} in the gravitational radiation.  In fact, we shall see below (section X) that this is also true of the {\em other,} ``pure spin,'' terms when there is a supertranslation mismatch.

\subsection{Minkowskian and strongly Minokwskian regimes}

In a Minkowskian regime $\RR$, one can solve the equation $\eth^2\lambda =\sigma$ to find a $u$-independent real potential $\lambda$ (and one could take its $j=0$ and $j=1$ parts to vanish, if desired).  Then if a cut $u=\lambda +\tau$, where $\tau$ is a translation, lies in $\RR$, it will be a good cut.  It is commonly asserted that in a Minkowskian regime one can always find good cuts, but this is not true.

For a counterexample, suppose $\lambda$ is $A\sin (m\phi)$ around the equator, with $m\geq 2$.  Then the mean value of $(\lambda +\tau)^2$ over the equator will be $\geq A^2/2$ for any translation $\tau$ (since $\tau$ can have only $m=0$ and $m=1$ components, which will, as functions on the equator, be $L^2$-orthogonal to $\lambda$), and hence $|\lambda +\tau|$ must be at least $|A|/\sqrt{2}$ at somewhere on the equator.  If ${\mathcal R}$ lies within the strip $|u|<|A|/\sqrt{2}$, then, no good cut can lie inside it.

If the Minkowskian regime persists for a sufficiently long interval, it will contain good cuts, so for example a half-infinite in- or out-regime will have this property:  but in general being able to find these cuts is an extra requirement.
I will say a regime is {\em strongly Minkowksian} if we can find a four-dimensional family of good cuts in it.  
Of course, one can always mathematically extend a Minkowskian regime to a strongly Minkowskian one, simply by replacing the shear on as large a domain as desired with the $u$-independent Minkowskian values.  

The distinction between Minkowskian and strongly Minkowskian regimes will
not matter for twistor theory --- twistor theory in effect automatically makes use of the strongly Minkowskian extension in any case.
For BMS-based approaches, the significance is mixed.  The approaches themselves, by their nature, do not single out any class of cuts, good or otherwise.  On the other hand, the interpretation of the BMS angular momenta at bad passive cuts is ill-understood.  It is conceptually helpful to put in ``by hand'' the strongly Minkowskian extension and work with the resulting good cuts.

\subsection{Minkowski space model for a Minkowskian regime}

In a 
Minkowskian regime ${\mathcal R}$, there is no gravitational radiation, and 
this turns out to mean the twistor equation will be integrable to fields $\omega^A=\omega^0\omicron^A +\omega^1\iota^A$ independent of the slicing.  
(This can also be verified from eqs. (\ref{omz}), (\ref{omone}).)
Then the twistor space $\T({\mathcal R})$ has canonically a vector-space structure.  
There is no magnetic shear, so the reality structure $\Phi$ arises as a pseudo-Hermitian form on $\T({\mathcal R})$ and there is an associated Minkowski space $\MM ({\mathcal R})$ which may be used for describing the energy--momentum and angular momentum.

The kinematic twistor $A_\z (Z )$ is defined (eq. (\ref{kintwist})) by integrating the two-form
\begin{eqnarray}
  \psi _{ABCD}\omega^C{\acute\omega}^D\epsilon_{A'B'}
\end{eqnarray}
over $\z$.  In the case of $u$-independent shear, this is
\begin{eqnarray}
  -2[ \psi _1(\omega^0)^2
   +  2\psi _2 \omega^0\omega^1 ] &{\overline m}_{[a}m_{b]}
  +2\psi_2 (\omega^0)^2 &l_{[a} m_{b]}\, ,\qquad
\end{eqnarray}  
and a direct computation shows that this is closed on $\scrif$.  {\em Therefore the angular momentum $A_\z (Z)$ is independent of the (active) cut $\z$ within ${\mathcal R}$.}  Certainly the angular momentum depends on the (passive) choice of origin --- but that choice is coded in the choice of the twistor $Z$, {\em not} the cut at which the integral is done.
In particular, the cut need not be a good cut, and there is no requirement the regime be strongly Minkowskian.

To interpret this more fully, we must look at how the Minkowski space $\MM ({\mathcal R})$ is defined.  If the regime ${\mathcal R}$ were strongly Minkowskian, the points in the Minkowski space would be the shear free cuts in ${\mathcal R}$.  However, even if the regime itself is only 
Minkowskian, the points in $\MM({\mathcal R})$ will be the the shear-free cuts of the mathematically fictive strongly Minkowskian extension.
The result of this is that, twistorially, the cut $\z$ appears simply as if it were a bad cut in a strongly Minkowskian regime; the twistorial construction in effect automatically creates this regime with its good cuts, and uses those as the origins.  

The argument just given is rigorous but abstract.  I will next show explicitly how the details work out for good passive cuts (but where the active cut is arbitrary), that is, how a special-relativistic angular momentum and energy--momentum, the appropriate transformation law, are recovered.  Then I will examine the angular momentum at an arbitrary passive cut; it is there we will find the change-of-origin formula interpreting supertranslations as direction-dependent translations.

\subsection{Twistors orthogonal to a cut}

Suppose $u=\z$ is a cut of $\scrif$; we should like to use twistor structures, as well as we can, to interpret this cut as a passive origin.  
That is, we should like to see if we can define an angular momentum.
If $\z$ were a good cut in Minkowski space, then it would be the intersection of the future light-cone of a point $p$ with $\scrif$, and the twistors through $p$, each one of the form $(ip^{AA'}\pi_{A'},\pi_{A'})$, would determine the components $\mu^{A'B'}\pi_{A'}\pi_{B'}$ of the angular momentum with origin $p$.

In a more general space--time, even if $\z$ is a good cut, there is no guarantee that it actually is the intersection of the future light-cone of any point with $\scrif$.  Nevertheless, there is a natural geometric property of the Minkowskian case we can take over:  we can choose the twistors to meet the cut orthogonally.  This can be done whether $\z$ is good or not.  We find here the twistor $Z(\z,\bm{\pi_{A'}})$ orthogonal to $\z$ and with projection $\bm{\pi_{A'}}$ to the base space.

We will suppose the field $\omega^0$ determining $\bm{\pi_{A'}}\in\Spin_{A'}$ has been fixed.
Recall from section VA that the twistors have the form
\begin{eqnarray}\label{terq}
  \omega^1 &=& \omega^0\eth\lambda -\lambda\eth\omega^0 +\alpha(u)\eth\omega^0
    +\beta (u)\overline{\omega^0}\, .
\end{eqnarray}
For a twistor to lie on the 
cut $u=\z$, we must have 
\begin{eqnarray}\label{terqal}
  \alpha (\z(\gamma))&=& \lambda (\z(\gamma ),\gamma )\, .
\end{eqnarray}
The null tangent to the cut is ${\overline m}^a+(\eth'\z) n^a$, and we require the twistor's tangent spinor
\begin{eqnarray}
\pi_{A'}\Bigr|_\gamma &=&\pi^{0'}\omicron_{A'}+\pi^{1'}\iota_{A'}\nonumber\\
  &=& \left(i(\eth\omega^0)\omicron_{A'} -i(\eth'\omega^1)\iota_{A'}\right)\Bigr|_\gamma
  \nonumber\\
  &=& \left(i(\eth\omega^0)\omicron_{A'} +i(\eth'\lambda\eth\omega^0
  -i\beta(\z(\gamma))\overline{\eth\omega^0})\iota_{A'}
  \right)\Bigr|_\gamma\qquad
\end{eqnarray}
be orthogonal to the null tangent. This gives the equation
\begin{eqnarray}
1:\eth'\z \Bigr|_\gamma&=& \left( \eth\omega^0: \left(\eth'\lambda\eth\omega^0 -
  \beta (\z(\gamma))\overline{\eth\omega^0}\right) \right) \Bigr|_\gamma\, ,
\end{eqnarray}  
where the colons indicate ratios (possibly infinite), or
\begin{eqnarray}\label{terqbe}
  \beta(\z(\gamma )) &=& -\left(\frac{\eth\omega^0}{\overline{\eth\omega^0}}\eth' (\z-\lambda)\right)
    \Bigr|_\gamma\, .
\end{eqnarray}        

In sum, given a cut $u=\z$, the twistor $Z(\z,\bm{\pi_{A'}})$ meeting $\z$ orthogonally at $\gamma (\bm{\pi_{A'}})$ is given by eqs. (\ref{terq}), (\ref{terqal}), (\ref{terqbe}).  (Recall that $\alpha (u)$ and $\beta(u)$ are determined from their values at any point on $\gamma$; eqs. (\ref{alphaeq}), (\ref{betaeq}).)

\subsection{Angular momentum at good passive cuts}

In our Minkowskian regime with $\sigma =\eth^2\lambda$, the good cuts have the form $u=\lambda +\tau$, where $\tau$ is a translation.
We may identify 
\begin{widetext}
\begin{eqnarray}
  A_{u_0}(Z(\lambda,\bm{\pi_{A'}} )) &=&
\frac{-i}{4\pi G}\oint_{u_0}\left\{
  \psi_1 (\omega^0)^2 
  + 2\psi_2 \omega^0 (\omega^0\eth\lambda -\lambda\eth\omega^0
    +u_0\eth\omega^0)\right\}\\
  &=& 2i\bm{\mu^{A'B'}}(\lambda )\bm{\pi_{A'}\pi_{B'}}\\
\end{eqnarray}
as $2i$ times the angular momentum, relative to (the passive choice of) good cut $u=\lambda$.

We will now show that the angular momentum measured about the (passive choice of) good cut $u=\lambda +\tau$ differs from the above by the correct change-of-origin formula.  We have
\begin{eqnarray}\label{transeff}
A_{u_0}(Z(\lambda+\tau,\bm{\pi_{A'}})) 
    &=&A_{u_0}(Z(\lambda,\bm{\pi_{A'}} )) 
    +\frac{-2i}{4\pi G}\oint_{u_0} \left\{ -\psi_2\left( \tau(\gamma )\omega^0\eth\omega^0
      +\left( \frac{\eth\omega^0}{\overline{\eth\omega^0}}\eth'\tau\right)\Bigr|_\gamma
       \omega^0\overline{\omega^0}\right)\right\}\label{medq}
        \, .
\end{eqnarray}    
The expression
\begin{eqnarray}
 \tau(\gamma )\eth\omega^0
      +\left( \frac{\eth\omega^0}{\overline{\eth\omega^0}}\eth'\tau\right)\Bigr|_\gamma
       \overline{\omega^0}
\end{eqnarray}       
occurring in the integral above can be written as
\begin{eqnarray}
-i \bm{\tau^{CC'}\omicron_C\omicron_{C'}}  \bm{\iota^{B'}\pi_{B'}}   \bm{\iota^A}
  +i\bm{t^{1B'}\pi_{B'}} \bm{\tau^{CC'}\iota_C\omicron_{C'}}  \bm{\omicron^A}
   &=& i\bm{\iota^{B'}\pi_{B'}} \bm{\tau^{CC'}\omicron_{C'}}
     \left( -\bm{\omicron_C\iota^A}+\bm{\iota_C\omicron^A}\right)\\
     &=& -i\bm{\iota^{B'}\pi_{B'}} \bm{\tau^{AC'}\omicron_{C'}}\, .
\end{eqnarray}  
\end{widetext}
Thus the contribution from the integral in eq. (\ref{transeff}) is
\begin{eqnarray}
  -2i\bm{P_{AA'}\iota^{B'}\pi_{B'}} \bm{\tau^{AC'}\omicron_{C'}\pi^{A'}}
  &=&-2i\bm{P_{AA'}} \bm{\tau^{AC'}\pi_{C'}\pi^{A'}}\, ,\nonumber\\
  &&
\end{eqnarray}  
so we have
\begin{eqnarray}\label{amse}
A_{u_0}(Z({\lambda +\tau},\bm{\pi_{A'}})) &=& 2i\bm{\mu^{A'B'}}(Z(\lambda,\bm{\pi_{A'}}) )\bm{\pi_{A'}\pi_{B'}}
  \nonumber\\
  &&\quad
-2i\bm{P^{AA'}} \bm{\tau_A{}^{B'}\pi_{B'}\pi_{A'}}
\, ,\qquad
\end{eqnarray}
or equivalently
\begin{eqnarray}\label{strans}
\bm{\mu^{A'B'}}(\lambda +\tau)
&=& \bm{\mu^{A'B'}}(\lambda )
-\bm{P^{A(A'}} \bm{\tau_A{}^{B')}}
\, .
\end{eqnarray}
This is the correct special-relativistic form of the change-of-origin transformation law, and so we see explicitly how twistors reduce the angular momentum of a stationary regime to the special-relativistic case.
(The Dray--Steubel BMS charges give the same result in this case, if the passive cut is taken to be good.)

\subsection{Angular momentum at a general passive cut}

We now ask for the angular momentum at an arbitrary passive cut $\z$ in a Minkowskian regime.  As pointed out above, this may be taken as $A(Z(\z,\bm{\pi_{A'}}))$, where $Z(\z,\bm{\pi_{A'}})$ is the twistor with base $\bm{\pi_{A'}}$ and meeting $\z$ orthogonally.

The key point is that (given $\bm{\pi_{A'}}$), this orthogonality condition only depends on $\z$ to first order at $\gamma$.
Therefore for each $\gamma$, we may apply the previous subsection's formulas, with the translation $\tau$ (now $\tau(\gamma)$) 
chosen so that $\lambda (\gamma )+\tau(\gamma )=\z (\gamma )$ and $\eth'\tau =\eth' (\z -\lambda)$ at $\gamma$.  
These conditions are
\begin{eqnarray}
  \bm{\tau^{AA'}\overline{\pi}_A\pi_{A'}} &=& (\z (\gamma )-\lambda(\gamma))
            \bm{t^{AA'}\overline{\pi}_A\pi_{A'}}\\
  \bm{\tau^{AA'}t_A{}^{B'}\pi_{A'}\pi_{B'}} &=& 
    (\eth'(\z -\lambda))(\bm{\iota^{A'}\pi_{A'}})^2\, .
\end{eqnarray}
We may also write the last condition as     
\begin{eqnarray}
\bm{\tau^{AA'}t_A{}^{B'}\pi_{A'}\pi_{B'}} &=& (\bm{t^{BB'}{\overline\pi}_B\pi_{B'}})
   \bm{t_A{}^{A'}\pi_{A'}}
   \frac{\partial (\z-\lambda )}{\partial{\bm{{\overline\pi}_{A}}}}\, .\nonumber\\
   &&
\end{eqnarray}
The solution to these conditions has a natural and straightforward appearance, once we express the supertranslation $\z-\lambda$ in a form adapted to the asymptotic spin space.

When, in the Bondi--Sachs formalism, we represent a supertranslation by a function $\alpha (\theta ,\phi )$, that function is really the contraction of the corresponding $BMS$ vector field with the tetrad covector $l_a$, which is normalized by the condition $\bm{t^al_a}=1$.  It is convenient here to, in effect, drop this normalization condition by contracting instead with $\bm{\overline{\pi}_A\pi_{A'}}$.  To do this, extend
$\alpha (\theta,\phi )$ to a function of $\bm{{\overline{\pi}}_A\pi_{A'}}$ by setting
\begin{eqnarray}\label{homalph}
 \alpha(\bm{{\overline{\pi}}_A\pi_{A'}} ) = \alpha(\gamma ) (\bm{t^{AA'}}
   \bm{{\overline{\pi}}_A\pi_{A'}})\, .
\end{eqnarray}
(So if $\alpha(\gamma )$ were a translation $\alpha(\gamma ) = \bm{v^al_a}$, we would have
$\alpha (   \bm{{\overline{\pi}}_A\pi_{A'}} ) =\bm{v^{AA'}{\overline\pi}_A\pi_{A'}}$.)

This new function is then defined on the future null cone.  Its derivative is defined in all directions tangent to the cone, and hence the gradient
\begin{eqnarray}\label{aldef}
  \frac{\partial}{\partial\bm{{\overline\pi}_A\pi_{A'}}} \alpha
\end{eqnarray}
is well-defined modulo terms proportional to $\bm{{\overline\pi}^A\pi^{A'}}$.

After a little algebra, we find
\begin{eqnarray}\label{sutrans}
\bm{\tau^{AA'}} (\gamma)&=& 
      \frac{\partial\left( \z -\lambda\right)(\bm{{\overline{\pi}_A}\pi_{A'}})}{\partial\bm{{\overline\pi}_A\pi_{A'}}} 
      \nonumber\\
      &&
            +\text{ a multiple of }\bm{{\overline\pi}^A\pi^{A'}}\, .\label{tueq}
\end{eqnarray}         
This quantity can be viewed as a direction-dependent translation, the direction determined by $\bm{\pi_{A'}}$  (which determines $\gamma$), modulo, for each direction, translations in that direction.  (It is not necessary for us to restrict that freedom, but I will discuss doing so below.)

We then have
\begin{eqnarray}\label{amsp}
A_{u_0}(Z(\z,\bm{\pi_{A'}} )) &=&
  2i\bm{\mu^{A'B'}}(\lambda )\bm{\pi_{A'}\pi_{B'}}\nonumber\\
  &&\quad
-2i\bm{P^{AA'}} \bm{\tau_A{}^{B'}}(\gamma )\bm{\pi_{B'}\pi_{A'}}
\end{eqnarray}
(independent of the freedom in $\bm{\tau^{AA'}}$).
The interpretation of this is that the angular momentum about $\z$ appears to be what we would get by a $\gamma$-dependent translation $\bm{\tau^{AA'}}(\gamma )$
from the angular momentum associated with a good cut.  
We cannot in this case represent the angular momentum at $\z$ by simply a spinor $\bm{\mu_{A'B'}}(\z )$, for the right-hand side of eq. (\ref{amsp}) has a more complicated dependence on $\bm{\pi_{A'}}$.  

The formula (\ref{amsp}) (together with the definitions (\ref{aldef}), (\ref{sutrans})) is the main result of this subsection.
Its interpretation 
is that the angular momentum in a Minkowskian regime but at a bad passive cut acquires $j\geq 2$ components.  
That is, for a good cut, the angular momentum $A(Z(\z,\bm{\pi_{A'}} ))$, regarded as a function of $\bm{\pi_{A'}}$, lies in a pure $j=1$ complex representation of the Lorentz group, but for a general cut there are $j\geq 2$ contributions as well.\footnote{To see that there are no $j=0$ contributions, it is easiest to go into a Bondi frame aligned with the energy--momentum.}  
The $j\geq 2$ contributions here are an indication that the center of mass is supertranslated relative to $\z$; we will see this in more detail in section~VIII.

The way the supertranslation $\z-\lambda$ comes in is significant.  It is (as it would be for a translation in Minkowski space) this quantity's gradient which enters, and the form of the change-of-origin term is (for each fixed $\bm{\pi_{A'}}$) Minkowskian.  (By contrast, for BMS charges the effect of a supertranslation on the angular momentum involves an integral over all directions of the supermomentum against the action of the Lorentz generator on the supertranslation.)

A few remarks:

It is the formula (\ref{amsp}) which is most natural expression of the angular momentum from the present point of view, but one can ask how closely we may make this correspond to more familiar expressions, that is, whether we could define an angular momentum spinor at $\z$.
We could take
\begin{eqnarray}
\bm{\mu^{A'B'}}(Z(\z, \bm{\pi_{A'}} ) )&=&
  \bm{\mu^{A'B'}}(\lambda )
  -\bm{P^{A(A'}} \bm{\tau_A{}^{B')}}(\gamma )\, ,\qquad
\end{eqnarray}
and in this sense one has a close parallel to the special-relativistic formula (\ref{strans}), the ordinary translation being replaced by a direction-dependent one, and so the angular momentum spinor given a directional dependence as well.  But note that it is only the contraction $\bm{\mu^{A'B'}}(Z(\z, \bm{\pi_{A'}} ) )\bm{\pi_{A'}\pi_{B'}}$ which is really defined by the arguments above, and that $\bm{\mu^{A'B'}}(Z(\z, \bm{\pi_{A'}} ) )$ would itself be direction-dependent.

Finally, let me comment on the freedom in the choice of $\bm{\tau^{AA'}}(\gamma )$.  There is no obvious way to fix this which respects universality, but there is a natural BMS-invariant prescription:  as long as the Bondi--Sachs momentum is timelike, one requires
$\bm{P_{AA'}}
\bm{\tau^{AA'}}(\gamma )=0$.

\subsection{Comparison of Minkowskian regimes}

In practice, we often have a system which is 
Minkowskian in some regimes, but emits radiation in others, 
and we wish to compare the Minkowskian regimes' angular momenta.  
For instance --- and this is the most-discussed case --- we may have a system which emits radiation for only a finite interval, and want to compare the initial and final angular momenta.
(Of course, the assumption that the radiation is only for a finite interval is an idealization --- as indeed is the assumption that Minkowskian regimes exist.  Really these are short-hands for considering limiting arguments, and assuming that deviations from these idealizations are small.)

Each Minkowskian regime
regime $\RR_j$ will have an associated Minkowski space $\MM (\RR_j)$ and angular momentum spinor field $\bm{\mu^{A'B'}}_{\RR_j}$ defined on that space, but if there are supertranslation mismatches between the good cuts of the different regimes
there can be no satisfactory Poincar\'e identification of these Minkowski spaces.  How can we compare their angular momenta, and so meaningfully talk of the angular momentum emitted in gravitational radiation between 
Minkowskian regimes?

We may resolve this at one level by simply noting that the twistorial angular momenta $A_{\RR_j}(Z)$ are all functions on the same twistor space $\TT$, and so the differences $A_{\RR_k}(Z)-A_{\RR_j}(Z)$ are well-defined.  Such a quantity represents the angular momentum, taking as (passive) origin the twistor $Z$, emitted in passing from $\RR_k$ to $\RR_j$.  
This would be the most natural quantity to consider from a purely twistorial point of view.

However, to make contact with more familiar treatments, we would like to express the change in angular momentum in terms of the special-relativistic data on the Minkowski spaces $\MM(\RR_j )$, $\MM(\RR_k)$.  
This will code the same concept, but in a different way.  
We know that some non-Poincar\'e contribution must be involved.

We can read this off from the results above.
Let $\lambda_k$ be a real $u$-independent potential for the shear in the regime $\RR_k$. 
Referring all angular momenta to the regime $\RR_j$, we express them all in terms of a choice of (passive) twistor $Z(\lambda _j+\tau,\bm{\pi_{A'}})$.  We
will have
\begin{eqnarray}
 &&A_{\RR_k}(Z(\lambda _j+\tau,\bm{\pi_{A'}}))=
    \nonumber\\
   &&\quad 2i\left(\bm{\mu^{A'B'}}_{\RR_k}({\lambda_k})
      -\bm{P^{AA'}}_{\RR_k} 
      (\bm{\tau_A{}^{B'}}_{kj} + \bm{\tau_A{}^{B'}})\right)\bm{\pi_{A'}\pi_{B'}}
      \, ,\qquad\label{keq}
\end{eqnarray}
where
$\bm{\tau^{AA'}}_{kj}(\gamma)$ is the angle-dependent translation corresponding to the supertranslation 
$\lambda_k(\gamma ) -\lambda_j(\gamma )$,
as in eq. (\ref{tueq}).

We can thus think of the angular momentum of $\RR_k$ expressed in $\RR_j$ as encoded in 
\begin{eqnarray}\label{angkj}
&&(2i)^{-1} A_{\RR_k}(Z(\lambda_j+\tau),\bm{\pi_{A'}}) =\\
&&\quad\left( \bm{\mu^{A'B'}}_{\RR_k}({\lambda_k})
      -\bm{P^{AA'}}_{\RR_k}  \bm{\tau_A{}^{B'}}
      -\bm{P^{AA'}}_{\RR_k} \bm{\tau_A{}^{B'}}_{kj}(\gamma )\right) \bm{\pi_{A'}\pi_{B'}}
      \, ,\nonumber
\end{eqnarray}
where the last term is the non-Poincar\'e contribution involving the supertranslation of $\RR_k$ relative to $\RR_j$.  
If we want the difference in angular momentum in two of the regimes, but expressed relative to our (passive) choice $\RR_j$, we have simply to take the differences in the quantities (\ref{angkj}) for the two values of $k$.

There is an aspect of this which is worth pointing out.  Let us look at the angular momentum in $\RR_k$ minus that in our reference regime $\RR_j$, so this could in particular be the case of the total angular momentum emitted by a system (taking $\RR_k$ the distant future and $\RR_j$ the distant past at $\scrif$).  The difference in the angular momenta is
\begin{eqnarray}\label{diffA}
(2i)^{-1}\left( A_{\RR_k}(Z(\lambda_j+\tau),\bm{\pi_{A'}})
  -A_{\RR_j}(Z(\lambda_j+\tau),\bm{\pi_{A'}})\right)\, ,\qquad
\end{eqnarray}
which is not antisymmetric in $j$ and $k$.  In other words, the difference of the angular momentum in $\RR_k$ from that in $\RR_j$, expressed in terms of $\MM(\RR_j )$, is not minus the difference in the angular momentum in $\RR_j$ from that in $\RR_k$, expressed in terms of $\MM(\RR_k)$.  The issue is the failure of a common Minkowskian structure to exist.

The difference (\ref{diffA}) in angular momenta is explicitly
\begin{widetext}
\begin{eqnarray}\label{diffa}
\left[ \bm{\mu^{A'B'}}_{\RR_k}({\lambda_k}) - \bm{\mu^{A'B'}}_{\RR_j}({\lambda_j}) 
      -\left( \bm{P^{AA'}}_{\RR_k} -\bm{P^{AA'}}_{\RR_j}\right)  \bm{\tau_A{}^{B'}}
      -\bm{P^{AA'}}_{\RR_k} \bm{\tau_A{}^{B'}}_{kj}(\gamma )\right] \bm{\pi_{A'}\pi_{B'}}
      \, .
\end{eqnarray}
\end{widetext}
It is the last term in the square brackets, giving the direction-dependent translation, which fails to be skew in $j$ and $k$.  This term is also 
what has been called a {\em longitudinal} contribution, meaning it depends not just on the difference in energy-momenta between the two Minkowskian regimes, but the net energy--momentum.  Workers have considered the possibility of such terms, but have generally used approaches which do not produce them.  (See Ref. \cite{BP2013}, and references therein.)

One final comment:  The last term in the square brackets
will not generally vanish even if the regimes $\RR_j$ and $\RR_k$ are not relatively supertranslated, because $\lambda_k-\lambda _j$ may well be a non-trivial translation.  (For instance, this will happen if $\lambda_j$ and $\lambda_k$ are indvidually chosen to be translation-free in the frames defined by the energy--momenta in $\RR_j$ and $\RR_k$, because those frames will generally be different.)  But then this term will provide the usual change-of-origin contribution needed to compare the first two terms.

\section{Quotient Minkowski spaces}

Gravitational radiation makes it impossible to assign a consistent Minkowski structure throughout the asymptotic regime.  These difficulties are associated with trying to reconcile the behavior of the field in different asymptotic directions.
In subsequent sections, I will take up directly the question of interpreting this directional dependence.  Here, however, 
I point out that for each fixed direction for the components ---  fixed $\bm{\pi_{A'}}$ ---
one does get a Minkowski-like structure, valid for all cuts, even when gravitational radiation is present.
These results are limited,
but worth setting out.

For each generator $\gamma$ of $\scrif$, we may consider the space of null geodesics ending on $\gamma$.  There is a three-dimensional family of these, which may be identified with the quotient $\MM/\bm{{\overline\pi}_A\pi_{A'}}$, where $\bm{\pi_{A'}}$ is the spinor labeling the generator.  In twistor space $\TT$, this is the three-real-dimensional 
{\em affine} subspace of the real elements of the fiber over $\bm{\pi_{A'}}$, determined by requiring $\alpha (u_0)-\lambda(\gamma ,u_0)\in\R$ when the twistors are specified in the form (\ref{omone}).  That is, {\em there is a direct natural identification of $\MM/\bm{{\overline\pi}_A\pi_{A'}}$ with the real elements of the fiber over $\bm{\pi_{A'}}$.}

The angular momentum, at an active cut $u=u_0$ but referred to a (passive) such twistor $Z$ is given by $A_{u_0}(Z)/(2i)$.  A very slight extension of the computations of the previous section (taking into account the fact that in a non-stationary regime the combination $\psi_2+\sigma\dot{\overline\sigma}$, and not just $\psi_2$, appears in the kinematic twistor), shows that the effect of a translation by $\tau$ on $Z$ is to change the parameters $\alpha$ and $\beta$ specifying the twistor by
\begin{eqnarray}
  \Delta\alpha &=&-\tau(\gamma )\\
  \Delta\beta &=& -\frac{\eth\omega^0}{\overline{\eth\omega^0}}\eth'\tau\Bigr|_\gamma\, .
\end{eqnarray}
Denote this translated twistor $Z+\Delta Z$.  Then, again from trivial adaptations of the previous subsection, we find the special-relativistic change-of-origin formula:
\begin{eqnarray}
  A_{u_0}(Z+\Delta Z) &=&A_{u_0}(Z) -2i \bm{P^{AA'}\tau_A{}^{B'}\pi_{B'}\pi_{A'}}
    \, .\qquad
\end{eqnarray}

So as long as we keep $\bm{\pi_{A'}}$ fixed (up to a complex multiple), we have a special-relativistic theory of angular momentum in $\MM/\bm{{\overline\pi}^A\pi^{A'}}$, valid for all cuts.  It is only when we want to relate the angular momentum in one direction to that in another (in conventional terms, when we want to compare different components) that more exotic structures need to be brought in.

\section{Spin and center of mass}

In special relativity, the representation of the angular momentum as a valence-two tensor (or spinor) is adapted to bringing out its transformation properties as one passes from one frame to another.  But for a given massive system, there is a preferred frame, that determined by its energy--momentum, and to {\em interpret} the angular momentum one almost invariably passes to that frame.  Then the space--space components give the spin, and the time--space component the mass-moments (or equivalently the center of mass).  Mathematically, these are elements of two $j=1$ representations of the orthogonal group.

In this section, I will explain the twistor treatment of spin and mass-moments.  We will see that each of these becomes a function of angle which has $j\geq 1$ components, with the $j\geq 2$ contributions due directly to the angular potential $\lambda$ for the shear.
This has two important effects:  It gives the center of mass and spin geometric, and physically attractive, interpretations.  And it suggests that the we think of the ordinary ($j=1$) angular momentum
and the shear ($j\geq 2$) as two parts of a single unified concept, the general relativistic angular momentum.

\subsection{Special relativity}

In special relativity, it is possible to recover the angular momentum and mass-moments by looking at the twistorial quantity
\begin{eqnarray}
A_{\alpha\beta}Z^\alpha {\acute Z}^\beta /I_{\alpha\beta}Z^\alpha {\acute Z}^\beta
\end{eqnarray}
subject to the following constraints:  the twistors $Z^\alpha$ and ${\acute Z}^\alpha$ are real, and also satisfy $A_{\alpha\beta}Z^\alpha Z^\beta = 0 =A_{\alpha\beta}{\acute Z}^\alpha {\acute Z}^\beta$. (We will see shortly that the parts $\pi_{A'}$, ${\acute\pi}_{A'}$ can be chosen arbitrarily, and there will be a further one-real-dimensional affine freedom for each twistor.)  With these assumptions, a straightforward calculation verifies the identity
\begin{widetext}
\begin{eqnarray}\label{ktident}
&&\left( A_{\alpha\beta}Z^\alpha {\acute Z}^\beta /I_{\alpha\beta}Z^\alpha {\acute Z}^\beta
\right)
\left(P^{AA'}{\overline\pi}_A\pi_{A'} P^{BB'}{\acute{\overline\pi}}_B{\acute\pi}_\beta \right)
  \nonumber\\
  &&\quad =
  (1/2)
  \left[ (2iP^A{}_{C'}\mu^{A'C'}+i\alpha P_bP^b t^{AA'}){\overline\pi}_A\pi_{A'}
     (P^{BB'}{\acute{\overline\pi}}_B{\acute\pi}_{B'})
     -(2iP^A{}_{C'}\mu^{A'C'}+i\acute\alpha P_bP^b t^{AA'}){\acute{\overline\pi}}_A{\acute\pi}_{A'}  
     (P^{BB'}{\overline\pi}_B\pi_{B'})\right]
     ,\qquad
\end{eqnarray}    
for some real $\alpha$, $\acute\alpha$, encoding the freedom remaining in $Z^\alpha$, ${\acute Z}^\alpha$ after imposing the constraints; note that the second term in the square brackets is the same as the first, with accented and unaccented occurrences of $\alpha$ and $\pi_{A'}$ exchanged.  (The notations $\alpha$, $\acute\alpha$ fit with the use of these symbols for the general-relativistic twistors in section VA, at the cut corresponding to the origin.)
The quantity $2i P^A{}_{C'}\mu^{A'C'}$ occurring in on right is  $M(J^{AA'}+iK^{AA'})$, where $M$ is the mass and $J^a$ is the spin and $K^a$ is the mass dipole.

We see then that 
\begin{eqnarray}
&&\Re \left( A_{\alpha\beta}Z^\alpha {\acute Z}^\beta /I_{\alpha\beta}Z^\alpha {\acute Z}^\beta
\right)
\left(P^{AA'}{\overline\pi}_A\pi_{A'} P^{BB'}{\acute{\overline\pi}}_B{\acute\pi}_\beta \right)
 \nonumber\\
  &&\qquad\qquad 
  =(M/2)
    \left( (J^{AA'}{\overline\pi}_A\pi_{A'})(P^{BB'}{\acute{\overline\pi}}_B{\acute\pi}_{B'})- (J^{AA'}{\acute{\overline\pi}}_A{\acute\pi}_{A'})(P^{BB'}{\overline\pi}_A\pi_{A'})\right)
   \, .\qquad
\end{eqnarray}  
As we may choose 
$\pi_{A'}$, ${\acute\pi}_{A'}$ freely, this determines the spin (if the energy--momentum is known).

The imaginary part of the equation (\ref{ktident}) brings in origin-dependent quantities ($K^a$, $\alpha$, $\acute\alpha$).  We will want to look for a way of interpreting this which does not require reference to a space--time origin, but it will be helpful to understand the details of the relation first.
Choose the time-axis aligned with the energy--momentum, so $P_a=Mt_a$.  Then 
\begin{eqnarray}\label{imkt}
&&\Im \left( A_{\alpha\beta}Z^\alpha {\acute Z}^\beta /I_{\alpha\beta}Z^\alpha {\acute Z}^\beta
\right)
\left(P^{AA'}{\overline\pi}_A\pi_{A'} P^{BB'}{\acute{\overline\pi}}_B{\acute\pi}_\beta \right)
\nonumber\\
  &&\qquad\qquad = (M/2)\left(
  (K^{AA'}+\alpha P^{AA'}){\overline\pi}_A\pi_{A'} (P^{BB'}{\acute{\overline\pi}}_B{\acute\pi}_{B'})
  -(K^{AA'}+\acute\alpha P^{AA'}){\acute{\overline\pi}}_A{\acute\pi}_{A'} (P^{BB'}{{\overline\pi}}_B{\pi}_{B'})\right)\, .
\end{eqnarray}  
\end{widetext}
We see from this that, if we are willing to make use of the origin to identify $\alpha$, $\acute\alpha$, then we may read off the mass-moment (for instance, by setting $\alpha=\acute\alpha =0$ and varying $\pi_{A'}$, ${\acute\pi}_{A'}$).

Suppose we restrict the twistors to pass through the point $x^a$.  Then we will have $\alpha = -x^{AA'}{\overline\pi}_A\pi_{A'}/(t^{AA'}{\overline\pi}_A\pi_{A'})$ (and similarly for $\acute\alpha$).  Note that if we took $x^a=K^a/M$, we will have 
$\alpha = -K^{AA'}{\overline\pi}_A\pi_{A'}/P^{AA'}{\overline\pi}_A\pi_{A'}$ and hence
\begin{eqnarray}\label{preveq}
  (K^{AA'}+\alpha P^{AA'}){{\overline\pi}}_A{\pi}_{A'} 
   &=&0\, .
\end{eqnarray}
This is close to saying that the expression (\ref{imkt}) will vanish on the center of mass.  It is not quite the same, because the center of mass for a special-relativistic system is not a point, but a world-line (with tangent $P^a$).  We may implement this by requiring, not eq. (\ref{preveq}), but
\begin{eqnarray}
  (K^{AA'}+\alpha P^{AA'}){{\overline\pi}}_A{\pi}_{A'} 
   &=&\left(\const \right) P^{AA'}{\overline\pi}_A\pi_{A'}\, ,
\end{eqnarray}
or equivalently, the left-hand side is a pure $j=0$ term in the center-of-momentum frame.

\subsection{General relativity}

Can we take the twistor constructions over to general relativity?  We may form the quantity 
$A(Z,\acute Z)/I(Z,\acute Z)$, 
and we may impose the constraints ($A(Z,Z) =A(\acute Z ,\acute Z ) =0$ and $Z$, $\acute Z$ real), but the constraints are significantly more complicated, because of the strongly nonlinear reality structure.  For this reason, it is not obvious
that we will have a parallel of the right-hand side of eq. (\ref{ktident}), as a difference in two complementary terms, each a product of a quantity depending on $\bm{\pi_{A'}}$ and one depending on $\bm{{\acute\pi}_{A'}}$.  Remarkably, it turns out this is possible.

The details of the algebra are straightforward but lengthy, and I will just outline the steps and give the result.  For twistors $Z$, $\acute Z$ in our form (\ref{omone}), 
write the kinematic twistor as      
\begin{widetext}
\begin{eqnarray}\label{ktb}
 A(Z,\acute Z ) &=& 2i\bm{\mu^{A'B'}\pi_{A'}{\acute\pi}_{B'}}
   +i\bm{P^{AA'}t_A{}^{B'}}(\alpha \bm{{\acute\pi}_{A'}\pi_{B'}}
   +\acute\alpha \bm{{\pi}_{A'}{\acute\pi}_{B'}} )
   +i\bm{P^{AA'}}(\acute\beta \bm{\pi_{A'}{\acute{\overline\pi}}_A}
     + \beta \bm{{\acute\pi}_{A'}{\overline\pi}_A} )\, ,
\end{eqnarray}
where $\bm{P^{AA'}}$ is the Bondi--Sachs energy--momentum as before, and
\begin{eqnarray}\label{ktc}
2i\bm{\mu^{A'B'}\pi_{A'}{\acute\pi}_{B'}}
  &=&\frac{-i}{4\pi G}\oint\{ \psi_1 \omega^0{\acute\omega}^0 +
  (\psi _2 +\sigma\dot{\overline\sigma}) 
  [\omega^0({\acute\omega}^0\eth\lambda-\lambda\eth{\acute\omega}^0
        )
    +{\acute\omega}^0({\omega}^0\eth\lambda-\lambda\eth{\omega}^0
      )]\}
       \, .
\end{eqnarray}
\end{widetext}
(The individual terms in eq. (\ref{ktb}) are {\em not} Lorentz-invariant, but depend on the Bondi frame.)

We use eq. (\ref{ktb}) to impose the constraints $A(Z,Z) =0$, $A(\acute Z ,\acute Z) =0$.  
We get
\begin{eqnarray}
  &&    2i\bm{\mu^{A'B'}\pi_{A'}\pi_{B'}}+2i\alpha\bm{P^{AA'}t_A{}^{B'}\pi_{A'}\pi_{B'}}
     \nonumber\\
     &&\quad
        +2i\beta\bm{P^{AA'}{\overline\pi}_A\pi_{A'}} = 0\, ,
\end{eqnarray}
and similarly for $\acute{Z}$.  We use these relations to eliminate $\beta$, $\acute\beta$.
The requirement that the twistors $Z$, $\acute{Z}$ be real, which is where the strongly nonlinear structure enters, is that $\alpha-\lambda (\gamma )=\alpha -\lambda (\bm{\pi_{A'}})$, $\acute\alpha -\lambda (\acute\gamma )=\acute\alpha -\lambda (\bm{{\acute\pi}_{A'}})$ be real.

Using these results, it is straightforward if lengthy to compute the quantity (\ref{ktident})
in the general-relativistic case.
Choosing the Bondi frame to be aligned with the energy--momentum, the result is
\begin{eqnarray}\label{almost}
&&\left( A(Z,\acute{Z})/I(Z,\acute{Z})\right)
  \left( \bm{P^{AA'}{\overline\pi}_A\pi_{A'}}\right)
  \left( \bm{P^{AA'}\acute{{\overline\pi}}_A{\acute\pi}_{A'}}\right)\nonumber\\
&&\quad= i\bm{\mu^{A'C'}P^A{}_{C'}{\overline\pi}_A\pi_{A'}P^{BB'}{\acute{\overline\pi}}_B{\acute\pi_{B'}}} \nonumber\\
&&\qquad
+(i/2)\alpha M\bm{P^{AA'}{\overline\pi}_A\pi_{A'}P^{BB'}{\acute{\overline\pi}}_B{\acute\pi}_{B'}}\nonumber\\
&&
  \qquad -\text{ the same with accented and unaccented}\nonumber\\
  &&\quad\qquad\text{ twistor quantities exchanged.}
\end{eqnarray}  
(The twistor quantities are $\bm{\pi_{A'}}$, $\bm{{\overline\pi}_A}$, $\alpha$, and the accented versions of these.)
We will look separately at the real and imaginary parts of this.

\subsubsection{Spin}

Comparing the real parts of eq. (\ref{ktident}) and (\ref{almost}), and remembering that the $\alpha-\lambda$ is real, we identify the spin as\footnote{Unfortunately, in Ref. \cite{ADH2007}, the sign of the $\Im\lambda$ contribution is given incorrectly in the corresponding result, its eq. (21), and this error is carried over to its eqs. (23), (24) and (1).}
\begin{eqnarray}\label{spin}
  &&\bm{J}(\bm{{\overline\pi}_A\pi_{A'}})\\
    &&\quad =-2\Im\bm{\mu^{A'C'}t^A{}_{C'}{\overline\pi}_A\pi_{A'}}
      -\Im (\lambda(\bm{\pi_{A'}}))\bm{P^{AA'}{\overline\pi}_A\pi_{A'}}\, .\nonumber
\end{eqnarray}
(Recall the frame is aligned with the energy--momentum, so $\bm{P^{AA'}}=M\bm{t^{AA'}}$.)
In special relativity, this would be a pure $j=1$ quantity $J^{AA'}{\overline\pi}_A\pi_{A'}$ (the first term on the lower line), but in general relativity there are $j\geq 2$ contributions from the second term, coming precisely from the magnetic part of the shear.  In fact, this formula identifies the potential $-\Im\lambda$ for the magnetic shear as the $j\geq 2$ part of the specific (that is, per unit mass) angular momentum.\footnote{Any $j=0$ or $j=1$ terms included in $\Im\lambda$ cancel out in the formula (\ref{spin}), using eq. (\ref{ktb}).}. This formula is an exact counterpart, in general relativity, of the Newman--Winicour interpretation of special-relativistic spin as a displacement of the center of mass into the complex.  Here the form of the spin corresponds to an imaginary supertranslation by $\Im\lambda$.

Note that the expression $J^{AA'}{\overline\pi}_A\pi_{A'}$, which eq. (\ref{spin}) generalizes, may be called the relativistic component of the spin determined by ${\overline\pi}^A\pi^{A'}$, but on account of the signature of the metric it is {\em minus} the usual definition of the component in the corresponding spatial direction.  One should correspondingly regard $\bm{J({\overline\pi}_A\pi_{A'})}$ as the relativistic quantity, but minus the spin associated with the spatial direction determined by $\bm{{\overline\pi}^A\pi^{A'}}$.  (In this connection, note that $\Im\lambda$ may have components of both parities, but the sign issue here is only the overall one for $\bm{J}(\bm{{\overline\pi}_A\pi_{A'}})$.)

\subsubsection{Center of mass}

Now let us look at the imaginary parts of eq. (\ref{ktident}) and (\ref{almost}).  Here $\alpha$, $\acute\alpha$ will enter.  We find
\begin{eqnarray}
&&\left(  \Re\bm{\mu^{A'C'}t^A{}_{C'}}+(\Re\alpha /2)\bm{P^{AA'}}\right)
  \bm{{\overline\pi}_A\pi_{A'}} \left(\bm{P^{BB'}{\overline\pi}_B\pi_{B'}}\right)
  \nonumber\\
&&
  \quad -\text{ the same with accented and unaccented}\nonumber\\
  &&\qquad\text{ twistor quantities exchanged.}
\end{eqnarray}  

Following the argument at the end of the previous subsection, we will
now aim to define the center of mass by
choosing $\alpha$ as a function of $\bm{\pi_{A'}}$ so that
\begin{eqnarray}
  \left(  2\Re\bm{\mu^{A'}{}_{C'}t^{AC'}}+(\Re\alpha )\bm{P^{AA'}}\right) \bm{{\overline\pi}_A\pi_{A'}}
\end{eqnarray}
is a pure $j=0$ quantity (in our center-of-momentum frame).  
We will have 
\begin{eqnarray}
  \Re\alpha &=&\left(\Re\alpha\right)_{\rm cm,\, 0}
    +\left(\Re\alpha\right)_{\rm cm,\, 1}
\end{eqnarray}
as a sum of $j=0$ and $j=1$ terms, with the first arbitrary and    
\begin{eqnarray}
  \left(\Re\alpha\right)_{\rm cm,\, 1} &=&-2\frac{\Re\bm{\mu^{A'C'}t^A{}_{C'}{\overline\pi}_A\pi_{A'}}}{
          \bm{P^{AA'}{\overline\pi}_A\pi_{A'}}}\, .
\end{eqnarray}
We recall (from section VIC) that each value of $\alpha({\bm{\pi_{A'}}})$ determines the point on the generator at which the real twistor strikes $\scrif$; this defines a one-parameter family of cuts
\begin{eqnarray}\label{zcm}
 \z &=& u_0+\lambda(u_0,\gamma ) -\alpha\nonumber\\
     &=& u_0 +  \Re\lambda(u_0,\gamma ) 
     -\left(\Re\alpha\right)_{\rm cm,\, 1}
    -\left(\Re\alpha\right)_{\rm cm,\, 0}
      \, ,\qquad
\end{eqnarray}
Where $u=u_0$ is the active cut.

The formula (\ref{zcm}) is a main result, giving the center of mass of the system as a one-parameter family of cuts, indexed by $\left(\Re\alpha\right)_{\rm cm,\, 0}$.  Each of the cuts is supertranslated relative to the Bondi system by $\Re\lambda$, which has the effect of canceling any ``bad gauge'' choice in the active cut $\z =u_0$.  
(For instance, in a Minkowskian regime, this construction automatically selects good cuts as the center-of-mass world-line, even if the active cut is bad.)
There is also a $j=1$ contribution $\left(\Re\alpha\right)_{\rm cm,\, 1}$, which may be regarded as fixing the conventional, three-dimensional, information in the center of mass, once the gauge issues associated with the electric shear have been taken care of.  Since relativistically the center of mass is a world-line, the parameter $\left(\Re\alpha\right)_{\rm cm,\, 0}$ is arbitrary.\footnote{The freedom to add $j=1$ terms to $\lambda$ cancels in eq. (\ref{zcm}), using eq. (\ref{ktc}).  The freedom to add $j=0$ terms is absorbed by $\left(\Re\alpha\right)_{\rm cm,\, 0}$.}
That we get a well-defined world-line, and that it is insensitive to what may be considered gauge perturbations of the active cut, is physically satisfying, and arguably compelling.  
(Compare \cite{ADH2021}.)

The center-of-mass formula (\ref{zcm})
is parallel to the one (\ref{spin}) for the spin.  Besides a conventional ($j=1$) term, there are $j\geq 2$ terms, coming from the shear.  The center of mass is supertranslated by $\Re\lambda$ (plus $j=0,\, 1$ terms) from the active cut; for the spin, there was a formal supertranslation by $i\Im\lambda$ (plus $j=0,\, 1$ terms).

One can choose a specific cut on the center of mass which is closest (in the sense of $L^2$ functions on the sphere) to the active cut $\z=u_1$ by requiring the $j=0$ part of 
$\Re\lambda(u_0,\gamma )      -\left(\Re\alpha\right)_{\rm cm,\, 1}   -\left(\Re\alpha\right)_{\rm cm,\, 0}$ to vanish.  Then
\begin{eqnarray}
  \bm{K}(\bm{{\overline\pi}_A\pi_{A'}})
    &=& M\left( \z -u_0\right) \nonumber\\
    &=&\left( \Re\lambda(u_0,\gamma )      -\left(\Re\alpha\right)_{\rm cm,\, 1}   -\left(\Re\alpha\right)_{\rm cm,\, 0}\right)\qquad
\end{eqnarray}
may be interpreted as the mass-moment with respect to the cut $\z=u_1$, again parallel to the spin formula (\ref{spin}).

\subsection{Angular momentum and shear}

In special relativity, we may have a number of systems, each with an energy--momentum and angular momentum represented by a covector and a tensor field on Minkowski space.  We compare them straightforwardly in terms of those covariant quantities, but each individually has a preferred timelike direction (along its energy--momentum), relative to which the spin and center of mass can be read off.

In general relativity, the covariant quantities are the angular momentum twistors $A_\z(Z)$ at the different cuts of $\scrif$; these are straightforwardly compared.
At each of these cuts, relative to the energy--momentum
we may extract the spin and center of mass.  
This determination is more involved than in special relativity, reflecting the physical reality that the system's natural center of mass may be supertranslated relative to $\z$ (and the corresponding Newman--Winicour interpretation of the spin).

Mathematically, although $A_\z(Z)$ is a quadratic form on each vector space $\T (\z )$, that vector space depends on the shear (through eq. (\ref{omone})), and when we isolate the spin and center of mass that dependence comes in explicitly.

Remarkably, we find that the spin and center of mass comprise not just the $j=1$ quantities familiar from special relativity, but also $j\geq 2$ terms, containing the information in the shear.  In other words, general-relativistically, the angular momentum is to be understood as comprising both the $j=1$ terms and the shear.

\section{Gravitational radiation and emission of angular momentum}

A central result of the Bondi--Sachs theory is that the energy--momentum in gravitational waves is purely quadratic in the radiation $\dot\sigma$ --- indeed, up to a factor, one simply projects the $j=0$ and $j=1$ components of $|\dot\sigma|^2$.  That this is second-order is a main reason that energy--momentum loss by radiation is small in most cases.  It is natural to ask what the corresponding results for angular momentum are.

The very simple form of the Bondi--Sachs expression is possible because both the emitted power and $\dot\sigma$ are dimensionless (in general relativity).  Torque, by contrast, has dimension mass (or length), and so something must set the scale of angular momentum emission in a radiating system.  The two natural quantities with this dimension are $\psi_2$ and $\sigma$, and we will find that they both enter.  

Two observations are worth making at this point.  The first is that $\sigma$ has a gauge character, and we must take this into account in understanding the physical degrees of freedom.  The second is that because the two dimensionful quantities $\psi_2$, $\sigma$ can enter, as well as the dimensionless $\dot\sigma$, the question of which (if any) contributions dominate in given circumstances may be involved.  This will be important.
(In the BMS-based approaches, the only dimensionful quantity to enter is $\sigma$, and the behavior is more uniform.)

There is a further point to bear in mind, a freedom in splitting radiative from non-radiative terms.  Because the emitted angular momentum will depend (in general) on $\psi_2$ and $\sigma$, and those quantities are in turn determined by (say) their initial values at some cut as well as $\dot\sigma$, the details of the formulas in general will depend on which initial cut is chosen.  
Those details will not affect the main points of the discussion, however.

With this preamble, the properties of the emitted angular momentum can be outlined.

The change in angular momentum contains terms which are first-, second- and third-order in the gravitational radiation.  This can be see by inspection of the twistor angular momentum formula (\ref{kintwist}), taking into account the $u$-dependence of the twistor field $\omega^1$ (eqs. (\ref{omone}), (\ref{alphaeq}), (\ref{betaeq})) and the evolution equations
\begin{eqnarray}
 {\dot\psi}_2 &=&-\eth^2\dot{\overline\sigma}-\sigma\ddot{\overline\sigma}
    -4\pi GT_{111'1'}  \label{psitwoev}\\
{\dot\psi}_1 &=& \eth\psi_2 -2\sigma\eth\dot{\overline\sigma} -8\pi GT_{(01)1'1'}
  \label{psioneev}
\end{eqnarray}
(where $T_{111'1'}$, $T_{(01)1'1'}$ are the appropriately conformally rescaled components of the stress--energy for material radiation\footnote{By material radiation, I mean stress--energy which is carried off across $\scrif$.}).
If we consider changes between Minkowskian regimes, then the first- and third-order terms are both proportional to the supertranslation offset
between the initial and final regimes.  If there is no mismatch, then the emission is quadratic (and agrees with the BMS-based formula).

That first-order contributions exist suggests that in the weak-field limit angular momentum emission could be a more significant effect than energy--momentum emission.  We will see that this is true, and that in fact there is a sense in which the supertranslation mismatches are important carriers of angular momentum.  That third-order contributions exist suggests that there may be important strong-field deviations from the BMS-based formula.  We will see that, while the deviations can be important, there are also bounds on them.

These points are developed in the next sections.

\section{First-order emission effects}

I consider here the change in angular momentum between two Minkowskian regimes.  Without loss of generality, we may suppose $u=u_0$ is a cut in the first regime and $u=u_1$ is a cut in the second (extending, if necessary, the regimes to strongly Minkowskian ones).  I will also consider the values of $\psi _1$, $\psi_2$ and $\sigma$ at $u_0$ to be given initial data, their values elswhere determined by the evolution equations.

\subsection{The vacuum case}

Notice that, as we work to first order in $\dot\sigma$, there is no net change in the Bondi--Sachs energy--momentum.  A computation gives
\begin{eqnarray}\label{firsto}
  \Delta A(Z)\Bigr|_{\text{first-order}}
   &=& \frac{-i}{4\pi G}\oint 2\psi_2(u_0) \omega^0\Biggl[\omega^0\eth\Delta\lambda -
     \Delta\lambda\eth\omega^0\nonumber\\
     &&
     +(\Delta\lambda)\Bigr|_\gamma\eth\omega^0
     +\left(\frac{\eth\omega^0}{\overline{\eth\omega^0}}\eth'\Delta\lambda\right)\Bigr|_\gamma
     \overline{\omega^0}\Biggr]\, .\qquad
\end{eqnarray}     

The terms on the second line of eq. (\ref{firsto}) give a contribution corresponding to interpreting the supertranslation mismatch $\Delta\lambda$ as a direction-dependent translation, as in section VI.  They correspond to a change in center of mass with $j\geq 2$ components only (in the frame of the Bondi--Sachs energy--momentum).

The terms on the right-hand side first line of eq. (\ref{firsto}) are more interesting.  They give a pure $j=1$ change in spin,\footnote{Since $\psi_2(u_0)$ and $\Delta\lambda$ are real, and only the $\omega^0$ part of the twistor enters.} but that $j=1$ change is sensitive to the $j\geq 2$ terms in the mass-aspect $\psi_2(u_0)$.  
Since the mass aspect changes by terms which are first-order or higher in the gravitational radiation (assuming no material radiation is emitted), to the order of the present approximation it could be evaluated anywhere in the interval under consideration.

I would like to caution that there is no simple criterion for determining how the first-order contribution compares to the second-order ones.  Although at a formal level the first-order terms dominate in the weak-field limit,
if there is no supertranslation offset, the first-order contribution vanishes (as does the third-order one), and the twistor definition agrees with the BMS-based, second-order, one.

In the approximation where we keep only terms up to quadrupoles, a computation gives
\begin{eqnarray}\label{foq}
\Delta \bm{J^a}\Bigr|_{\text{first-order, quadrupole}}
 &=&\frac{4}{15G} \bm{\epsilon^{ab}{}_c\psi^{cd}}\Delta\bm{\lambda_{db}}
\end{eqnarray}
in standard three-tensor notation,
where $\bm{\psi_{ab}l^al^b}$ is the quadrupole contribution to $\psi_2$ and $\Delta\bm{\lambda_{ab}l^al^b}$ is the change in the quadrupole part of $\lambda$.
(Note that this is structurally similar to the Newtonian $\Delta{\bm{r}}\times\bm{p}$, with 
corresponding displacements
$\Delta{\bm{r}}$ and $\Delta\bm{\lambda_{ab}}$, and momentum or mass terms $\bm{p}$ and $\bm{\psi_{ab}}$ combined via a cross-product.)

I will close this subsection by giving a directly physical explanation for the $j=1$ part of the expression (\ref{firsto}).

Consider a system of small bodies, which initially are all freely falling and can be well-modeled over an interval of interest as following geodesics in Minkowski space --- that is, the mutual gravitational interactions of the bodies can be neglected.  The bodies may well be in relative motion, however.  If we wish to measure the spin angular momentum from asymptotic geometric data, we do so by working out the linearized gravitational response.  We set up a Bondi coordinate system, choosing the time-axis to be aligned with the total energy--momentum $\bm{P_a}$.  We also choose the Bondi system so that the $u=\const$ cuts are good, that is, we have $\sigma =0$ initially.  Then we compute
\begin{eqnarray}\label{simpj}
  2i\bm{\mu^{A'B'}\pi_{A'}\pi_{B'}} = \frac{-i}{4\pi G}
     \oint \psi_1 (\omega ^0)^2\, ,
\end{eqnarray}
as in section VI.  (We have $\lambda=0=\omega^1$.)

Now let us suppose one of these bodies fissions into two, which move off from each other with some relative velocities.  This fission gives rise to a brief burst of gravitational waves.  We will assume that $\dot\sigma$ is nevertheless very small.  Then we may neglect the energy radiated in the waves, relative to all the masses and kinetic energies which appear.  In particular, the Bondi--Sachs energy--momentum $\bm{P_a}$ will not change, to this approximation.

We want the spin angular momentum after the emission of radiation.  In this regime, the $u=\const$ cuts will now have a shear $\sigma$, purely electric and $u$-independent, so these will no longer be good cuts.  We may pass to the good Bondi coordinate $\acute{u}=u-\lambda$ (where $\eth^2\lambda =\sigma$), and compute the angular momentum in this new Bondi frame.
To do this, we find the curvature components (after the fission, and in the new frame). There are contributions from both the evolution of the system and the change in spin-frame (although some of these vanish).

We will have 
\begin{eqnarray}
  {\acute\psi}_2\Bigr|_{\text{after}} &=&\psi_2\Bigr|_\text{before}-\eth^2\overline\sigma
\end{eqnarray}
and 
\begin{eqnarray}
  {\acute\psi}_1\Bigr|_{\text{after}} 
    &=&\left(\psi_1 -\lambda{\dot\psi_1} -3(\eth\lambda)\psi_2\right)\Bigr|_{\text{before}}\\
  &=& \left( \psi_1-\lambda\eth\psi_2 -3(\eth\lambda)\psi_2\right)\Bigr|_{\text{before}} 
\end{eqnarray}
(to this order).
Thus the net change to $2i\bm{\mu^{A'B'}\pi_{A'}\pi_{B'}}$ will be
\begin{eqnarray}
&&2i\Delta\bm{\mu^{A'B'}\pi_{A'}\pi_{B'}}\nonumber\\
  &&\qquad =\frac{-i}{4\pi G} \oint \{ (-\lambda\eth\psi_2-3(\eth\lambda)\psi_2)(\omega^0)^2\}
  \nonumber\\
  &&\qquad=\frac{-i}{4\pi G} \oint \{ -2(\eth\lambda)\psi_2(\omega^0)^2 +2\lambda\psi_2
    \omega^0\eth\omega^0\}\, ,\qquad
\end{eqnarray}    
in agreement with the first line of eq. (\ref{firsto}).

In other words, the supertranslation mismatch gives rise to a change in the spin because of the need to choose a geometrically favored coordinate system (with good $u=\const$ cuts) to apply the simple formula (\ref{simpj}) for the spin.

It should be clear that this argument does not really depend on the special-relativistic system being composed of small bodies, and changing by fission.  What was actually used was just that one had a transition from one Minkowskian regime to another, that it was enough to work to first order in $\dot\sigma$, and that no material radiation escaped the system.

\subsection{Mixed matter-radiation contribution}

Material contributions do not appear explicitly in the twistor formula for $A_\z(Z)$, but the do enter when we compute differences $\Delta A(Z)$, through the evolution equations
(\ref{psitwoev}), (\ref{psioneev}) for $\psi_1$ and $\psi_2$.
If we, as before, regard initial data as given at $u=u_0$ and now both $\dot\sigma$ and the asymptotic values of the components of the stress--energy as determining the geometry elsewhere at $\scrif$, we find the additional material terms
\begin{eqnarray}\label{fomm}
\Delta A(Z)\Bigr|_{\text{matter}}
  &=&2i\oint\int_{u_0}^{u_1} \Bigl[ T_{(01)1'1'}(\omega ^0)^2 \\
  &&
   +T_{111'1'} \omega^0 \left(
    \alpha (u)\eth\omega^0 +\beta (u)\overline{\eth\omega^0}\right)
       \Bigr]\, du
       \, .\qquad\nonumber
\end{eqnarray}    

What is noteworthy here is that the second line of the formula (\ref{fomm}) depends on the gravitational radiation field, through $\alpha(u)$ and $\beta (u)$.  The terms coupling the material to the gravitational radiation are explicitly
\begin{eqnarray}\label{fmix}
&&\Delta A(Z)\Bigr|_{\text{mixed}} =2i\times\\
  &&\ \oint\int_{u_0}^{u_1} T_{111'1'} \omega^0 \left(
   \lambda (u,\gamma)\eth\omega^0 
   +\left(\frac{\eth\omega^0}{\overline{\eth\omega^0}}
     \eth'\lambda\right)\Bigr|_\gamma \overline{\omega^0}\right)\, du\, ,\qquad\nonumber
\end{eqnarray}   
where $\lambda$ is determined from $\dot\sigma$.  The terms in the expression (\ref{fmix}) are thus bilinear, in the stress--energy and the gravitational radiation.
Note that the expression is sensitive to the values of the potential $\lambda$ {\em throughout} the radiative interval, and not just to the supertranslation mismatch of the ends.  (In particular it will generally bring in the magnetic part of the shear.)
Its angular dependence may also be quite complicated, because of the highly nonlinear dependence of $\lambda$ on $\bm{\pi_{A'}}$.  
These points mean that in particular this term could produce $j=1$ contributions to both the spin and the center of mass (as well as $j\geq 2$ contributions).  

As was the case for the purely gravitational first-order term, there is no simple criterion for comparing the relative sizes of the mixing term (\ref{fmix}) with the quadratic effects.  Very roughly, the question would be how $G$ times what we may call the emitted material aspect $\int _{u_0}^{u_1}T_{111'1'}\, du$ compares with typical values of the shear within the period of gravitational radiation.

\section{Third-order effects}

The third-order contribution to the emitted angular momentum is
\begin{widetext}
\begin{eqnarray}\label{fotw}
\Delta A\Bigr|_{\rm third-order}&=& \frac{-2i}{4\pi G}     \oint \left(\int_{u_0}^{u_1} |\dot\sigma|^2\, du\right)
  \omega^0\left( \omega^0\eth\Delta\lambda -(\Delta\lambda -\Delta\lambda\Bigr|_\gamma)
    \eth\omega^0 +\left(\frac{\eth\omega^0}{\overline{\eth\omega^0}} \eth'\Delta\lambda
      \right)\Bigr|_\gamma\overline{\omega^0}\right)\, ,
\end{eqnarray}      
\end{widetext}
where $\Delta\lambda =\lambda (u_1)-\lambda (u_0)$.
This formula has a number of interesting properties: 
(a) It is explicitly proportional to what I have called the emitted energy aspect (the inner $u$-integral).
(b) It is also explicitly proportional to the difference $\Delta\lambda$ in the angular potential for the shears, which is the supertranslation mismatch between the bounding regimes.
(c) It depends only on the projection $\omega^0$ of the twistor to the base space, not on the twistor's position within the fibre.  
In this sense, the contribution is ``pure'' angular momentum.
(d) A little work shows the $j\geq 2$ contribution to be purely electric, that is, to affect the center of mass but not the spin.  (For the $j=1$ terms, in general the question of which are electric and which magnetic is frame-dependent.)

Combining points (a) and (b), we see that the third-order term is essentially bounded by the radiated energy times a pointwise supremum of a quantity constructed from $\Delta\lambda$ and $\eth\Delta\lambda$ (and $\omega^0$).  
The third-order term is formally close to the first-order one (\ref{firsto}), with the radiation mass-aspect replacing $\psi_2$, suggesting that in many cases it will be bounded by (roughly) the first-order term (\ref{firsto}), but, because of the different
angular dependences of the radiation mass-aspect and $\psi_2$ this is only an observation to start more careful analyses of particular cases from.

\section{Discussion}

When we attempt to treat angular momentum in general relativity, we seek to extend a familiar concept to a qualitatively new realm of physical conception.  We have no assurance that success is possible, and we cannot even be very precise about what success would mean.  We hope to uncover structure which will provide us with new physical insights; in particular, we hope to identify specially important quantities.

It is natural to approach this problem by asking what, at the deepest level we presently understand, seems to underly angular momentum, and conserved quantities generally.  This, though, leads immediately to a foundational conflict:  we are used to thinking of energy--momentum and angular momentum as conjugate to isometries of space--time, but it is precisely those invariances which are discarded as we pass to general relativity.

There are different reasonable responses to this.  One might seek formal structures in general relativity analogous to the isometries, and aim to base the treatment on those; we may regard the BMS charges as the result of such a program.  
But alternatively
we can take the conflict as a signal that we should reconsider our ideas about what the bases of the theory are.  The group-theoretic understanding, which we have supposed foundational, may rather be an especially beautiful specialization, to the Minkowski-space case, of some other, less obvious, structure existing in general relativity.  

\subsection{The twistor proposal}

Underlying the proposal developed here were two main ideas:  
that the most useful property of angular momentum is its {\em comparability} between different systems (or between different configurations of one system),
for this is what is needed to formulate a statement of conservation; and
that Penrose's 
remarkable quasilocal twistor construction seems to capture something deep about general-relativistic kinematics.
To meld
these, one looks for a {\em universal} twistor structure at $\scrif$, that is, a twistor space which is defined only in terms of the structure common to all Bondi--Sachs spaces.

There is a distinguished class of twistors, the {\em null} or {\em real} twistors, which can be identified with pairs $(\gamma ,\pi_{A'})$ of a null geodesic and a tangent spinor.  These are closely liked to angular momentum, for in special relativity the component
\begin{eqnarray}
  \mu^{A'B'}\pi_{A'}\pi_{B'}
\end{eqnarray}
of the angular momentum selected by $\pi_{A'}$ is independent of the choice of origin along $\gamma$.  This suggests viewing the angular momentum, not as a skew tensor-valued field $M_{ab}(x)$ on space--time points, but as a function $A(Z)$ on twistor space.  Each twistor carries some information about a space--time origin (the geodesic) and a choice of complex component (the spinor).

Making use of Penrose's quasilocal twistor construction, we can carry these ideas over to general relativity.  We can define a twistor space $\TT$, which is BMS-invariant, indeed universal for Bondi--Sachs space--times.  The angular momentum at any cut $\z$ is a function $A_\z(Z)$ on this space, and so angular momenta at different cuts are directly comparable.

Because the universal structure on $\TT$ is weaker than that in special relativity, we must explain how the function $A_\z(Z)$ is related to more familiar concepts, of the angular momentum as a tensor or spinor field, of spin and center of mass.  We know we cannot recover those structures unmodified, but we do want to be able to view their general-relativistic forms as physically comprehensible modifications of the special-relativistic case.

The simplest situations are the Minkowskian regimes.
In any such one $\RR$, the theory creates an associated Minkowski space $\MM (\RR )$, and the energy--momentum $\bm{P_a}$ and angular momentum $\bm{\mu_{A'B'}}$ can be viewed as those of a special-relativistic system on $\MM (\RR )$.  This space, $\bm{P_a}$ and $\bm{\mu_{A'B'}}$ are independent of the choice of active cut in $\RR$.  (In particular, the center-of-mass worldline is unambiguously defined.)

When we compare relatively supertranslated Minkowskian regimes, twistors give a formula which is an understandable extension of the special-relativistic one.  The effect of the supertranslation is to contribute a change-of-origin term which has the familiar algebraic cross-product structure, with one factor being the energy--momentum and the other a direction-dependent translation derived from the supertranslation.  This gives two, related, qualitatively new features.  First, the direction-dependence of the translations in the change-of-origin terms means that when we refer the angular momentum of one Minkowskian regime to the Minkowski structure of another, it acquires $j\geq 2$ terms.  Second, the difference in angular momenta has a ``longitudinal'' contribution proportional to the supertranslation mismatch.

At any cut (whether in a Minkowskian regime or not), we found that the spin and center of mass could be computed.  Each of these had a familiar $j=1$ contribution, but also $j\geq 2$ terms, coming from the angular potential for the Bondi shear.  For the center of mass, these extra terms gave a supertranslation which amounted to canceling what we would like to regard as any spurious contributions arising from a supertranslation of the active cut $\z$ relative to good cuts.  The $j\geq 2$ parts of the spin were interpretable as a supertranslation of the center of mass into the complex, in keeping with a special-relativistic observation of Newman and Winicour.  (For some more general comments on the connection between complex displacements and angular momentum, see Ref. \cite{Newman:2014}.)

This leads to a curious situation, where we have a strong physical argument for the correctness of the center of mass, and a strong mathematical coherence which makes us take the spin result seriously --- but we do not at this point have a full physical understanding of the spin.  

An especially interesting facet of this are its parity properties (mentioned at the end of Section VIII.B.1), which also underscore the difference between the twistor and the BMS approaches.
Let $\bm{\ell_{AA'}}=\bm{{\overline\pi}_A\pi_{A'}}$, and let $\bm{{\acute\ell}_{AA'}}$ be the null vector with spatial part reflected.   
Then we have in general
$\bm{J}(\bm{{\acute\ell}_a})\not= -\bm{J}(\bm{\ell_a})$, that is, 
the spin as a function of direction need not be parity-odd, because there may be parity-even parts of $\Im\lambda(\bm{\ell_a})$.  

This is arrestingly different from special relativity, and from the BMS-charge results.  There is no argument against it in the present context, however.  (The reason, usually, that the angular momentum is parity-odd is that it is taken to be conjugate to rotations, and a rotation in the positive sense about a direction is the same as the rotation in the negative sense about the opposite direction.  But here the direct link with rotations is not present.)  Moreover, the corresponding result for the center of mass (that the supertranslation $\Re\lambda$ may have parity-even parts) does not seem problematic at all.  Still, these comments show us only that there is no basic contradiction in the twistor spin having even-parity parts; they do not provide a physical elucidation of this possibility.

There were significant departures from the BMS-based formulas for the emission of angular momentum.  Whereas the BMS-based formulas were quadratic in the gravitational radiation, the twistor formulas had first- and third-order terms, present when there was a supertranslation offset of the shear in the period after the radiation to the period before; there was also an interesting mixed gravitational radiation--radiated stress--energy term.  One consequence of these formulas is that the first-order offsets in shear are consistently interpreted as exchanges of angular momentum with the gravitational radiation field.  This was discussed in some detail in the case of the first-order corrections derived from a special-relativistic system, and it was shown this measure of angular momentum arose in a physically natural way.

The twistor definition should not be regarded as divorced from, much less in opposition to, group-theoretic structure --- it is rather a question of which group comes in, and how.  One could say that in the twistor approach the Poincar\'e structure is minimally weakened in order to adapt to general relativity.  (This comes through in the existence of the quotient Minkowski spaces $\MM/\bm{{\overline\pi}_A\pi_{A'}}$, in the existence of twistor space $\TT$ as a manifold, and in way $A_\z(Z)$ comprises energy--momentum and angular momentum.)  By contrast, the BMS-based approaches make the transition to general relativity, not by what one can view as a single weakened Poincar\'e structure, but by introducing an infinite-dimensional family of Poincar\'e groups.

\subsection{Degrees of freedom}

It is worth understanding just what degrees of freedom are counted as angular momentum, and how these are related to the geometry.

Penrose showed that, for each active cut $\z$, the function $A_\z(Z)$ was determined by a ten-real-dimensional set of parameters, comprising the Bondi--Sachs energy--momentum and six further ones, which we interpret as angular momentum.  But this is relative to the linear structure $\T(\z)$, which depends on the shear.  In other words, the particular ten real degrees of freedom  depend on the cut chosen, through the shear.  In fact, this linear structure is equivalent to the shear (for one can recover $\sigma =(\eth\omega^1)/\omega^0$, independent of the twistor chosen).  Thus a fuller statement is that $A_\z(Z)$ codes ten real parameters, relative to knowledge of the shear.

When we evaluate, not just $A_\z(Z)$, but the spin and the center of mass, we impose reality conditions on the twistors, and the shear enters more explicitly:  in fact, all the infinite-dimensional degrees of freedom of the shear, at the active cut, are interpreted as parts of the general-relativistic angular momentum.  We have a strong motivation for accepting this interpretation, for it is what enables the twistor definition to compensate for ``bad'' contributions to the choice of active cut.

When we compare angular momenta at {\em different} cuts, we must take into account their differing shears.  The function $A_\z(Z)-A_{\acute\z}(Z)$ will not be quadratic with respect to either $\T(\z)$ or $\T({\acute\z})$, in general; it will be highly nonlinear.  It has a well-defined existence on the universal twistor space $\TT$, but the possible such functions form an infinite-dimensional family.

We may compare this with the BMS charges.  Each of these approaches brings in, in some sense, an infinite-dimensionality to the conserved quantities, even at a given active cut:  the twistors bring in the shear; the BMS charges the supermomenta (equivalently, the passive cuts).  The characters of these are rather different, in that the shear may be viewed as coding a conformally covariant part of the second fundamental form of the cut (that is, its first-order extrinsic geometry), whereas the supermomenta code the mass aspect (a second-order piece).  We have seen that $j\geq 2$ contributions to the twistor angular momentum come in precisely in the shear; for the BMS charges, for fixed $\z_{\rm act}$, $\z_{\rm pas}$, there are no $j\geq 2$ parts, but the effect of the different choices of $\z_{\rm pas}$ comes in through the $j\geq 2$ parts of the mass aspect.  (The twistors at any one cut only bring in a finite amount of information about the mass aspect.)

\subsection{Connection with canonical relativity}

Underlying the BMS-based approaches is the expectation of a link between conserved quantities and structure-preserving motions, and it is worth revisiting these points.
In general relativity, the link is expected to be provided by canonical (Hamiltonian or symplectic) mechanics.  For radiative problems, one would presumably phrase this in terms of data on what we may schematically designate an \fanm{}-shaped Cauchy surface (the middle portion lying in the physical space--time, and meeting a cut $\z$ of $\scrif$, and the legs of the \fanm\, representing the portion of $\scrif$ to the past of $\z$).  
From this perspective, it is natural to investigate BMS motions, and one would hope to be able to identify conjugate Hamiltonian functions, which might well lead to the BMS charges of Dray and Streubel.  (See Wald and Zoupas \cite{Wald_2000} for a suggested such identification, and Ref. \cite{ADH2021} for related comments.)

In this context, one seeks to understand a given space--time by considering the evolution of Cauchy surfaces, the BMS vector fields enter naturally to generate this,
and it may well turn out the BMS charges are central.  But what, then, would be the canonical interpretation of the twistor quantities?

This is an interesting question, and a possible avenue for gaining a deeper understanding.  The twistor angular momenta would presumably be well-defined functions on any properly defined phase space for Bondi--Sachs space--times.  One could take the angular momenta as Hamiltonian functions and, at least formally,\footnote{In these infinite-dimensional cases, there are sometimes technical obstructions to doing this.} solve for the associated vector fields on phase space, and Poisson brackets.  Because the twistor approach is BMS invariant, the vector fields and brackets should be so as well.\footnote{If, as one would hope, gauge issues involving changes of data on the part of the Cauchy surface in the finite space--time can be satisfactorily decoupled.}

The vector fields would represent perturbations of the space--time, preserving the component of the twistor angular momentum in question.  There would be no requirement, though, that those perturbations were induced by diffeomorphisms, and indeed (given that those which {\em are} induced by diffeomorphisms presumably come from the BMS approach) one would expect the contrary.  We would hope for insights by finding out just what the perturbations were.

These considerations suggest that the BMS charges and the twistor angular momenta may be best viewed not so much as competing proposals as answers to different questions --- aspects of angular momentum which would be equivalent in other contexts, but are distinct in general relativity.

\section*{Appendix:  Bondi--Sachs space--times}

Bondi--Sachs space--times model general-relativistic systems which can be considered isolated in the sense that one can give a sharp definition of gravitational (and other speed-of-light) radiation escaping from them.  This amounts to making hypotheses about certain features of their asymptotics.  In the original papers, these were formulated in terms of coordinate expressions of the metric and fall-off conditions.  Penrose showed that these could be recast in terms of the existence of an idealized boundary, future null infinity $\scrif$, for the space--times. 

Penrose's formalism is especially useful because it allows the asymptotic structure to be represented in a concise geometric way.  In fact, for the questions in this paper, the analysis is most clearly and conveniently done at $\scrif$ itself.
In this appendix, I will start by explaining and motivating the asymptotic geometry; the Bondi--Sachs space--times will be those which do admit such structures.  In fact, strictly speaking, Bondi and Sachs dealt with the vacuum case, but the analysis here allows some stress--energy (for example, such as might be expected from electromagnetic radiation), decaying as one approaches $\scrif$.
For technical details not given here, see Ref. \cite{PR1986}.

A physical Bondi--Sachs space--time $(\widehat{M} ,{\widehat g}_{ab})$ is required to embed as the interior of a manifold with boundary $M$ (with boundary $\scrif$).  The boundary is the zero-set of a non-negative suitably regular function $\Omega$ on $M$, with $\nabla_a\Omega$ nowhere zero on $\scrif$, and additionally 
there is a Lorentzian metric $g_{ab}$ on $M$ such that
$g_{ab}=\Omega^2{\widehat g}_{ab}$ on $\widehat{M}$.  (It is generally assumed that $M$ is $C^4$ in a neighborhood of $\scrif$ and that $g_{ab}$ and $\Omega$ are $C^3$.)

The boundary $\scrif$ is required to be made up of points which are the future limits of null geodesics in the physical space--time; for this reason it is called {\em future null infinity}.  (Not all null geodesics need have end-points on $\scrif$; there could, for example, be black holes.)  
Under reasonable hypotheses on the fall-off of the
physical stress--energy near $\scrif$, this boundary is necessarily null.  

That $\scrif$ is null means that we may pass to its quotient by its generators, and this quotient may be interpreted as the (two-dimensional) space of {\em asymptotic null directions}.  That such a well-defined space exists (for all the five-dimensional family of null geodesics meeting $\scrif$) is a central feature of the asymptotic structure.  We assume on physical grounds that $\scrif$ is diffeomorphic to $\R\times S^2$, with the $\R$ factors the null generators of $\scrif$.  (Under slightly stronger assumptions, this can be proved.)
Then the set of asymptotic null directions is diffeomorphic to $S^2$.


A key consequence of the Bondi--Sachs asymptotics is that the quotient $\scrif/\text{generators}$ is not just a sphere in the differential-topological sense, but has a well-defined {\em conformal}, or equivalently {\em complex}, structure.  (This too is due to the matter falling off as one approaches $\scrif$; that forces the spin-coefficient $\sigma '$, measuring the shear up the generators of $\scrif$, to vanish.)
In other words, the space of generators has naturally the structure of a Riemann sphere.
For this reason, the Newman--Penrose operators $\eth$, $\eth'$ (defined in terms of the complex structure)  are deeply bound with the Bondi--Sachs asymptotics.
The motions preserving this complex structure are the fractional linear transformations, isomorphic to the proper orthochronous Lorentz group.  It is this structure which underlies the existence of asymptotically constant spinors, vectors and tensors.

If we choose a unit sphere metric on this space of generators, it can be pulled back to $\scrif$ (as a tensor field --- it will be degenerate as a metric), and this fixes the freedom in the transverse derivative of $\Omega$ at $\scrif$, and 
this in turn can be used to set the scale of the the vector field $n^a$ tangent to the generators of $\scrif$.  
(In fact, then we have $n_a=-\nabla_a\Omega$ at $\scrif$.)
Although the particular scale will depend on the unit sphere metric chosen, different such choices will change $n^a$ by factors constant along each generator, and thus each generator acquires a well-defined affine structure.  A {\em Bondi parameter} (adapted to a particular unit-sphere metric) on $\scrif$ is a suitably smooth function $u$ with $n^a\nabla_a u =1$.

Without additional hypotheses, the generators need not be {\em infinitely long} (that is, the Bondi parameter need not take all real values).
The analysis of this paper does not require this condition, but for simplicity I have written in a few places as if the generators are infinitely long.

The foregoing sketch explains the basis for most of the analysis, that $\scrif$ is a bundle of affine lines (or segments) over $S^2$. 
To relate this to the original Bondi--Sachs construction, one extends the Bondi parameter $u$ into the physical space--time, by requiring $u$ be a null coordinate.  The angular coordinates in the finite space--time are then taken to be constant along the generators of the $u=\const$ hypersurfaces.  One may introduce an affine parameter $r$ along these generators, so that $\Omega\sim r^{-1}$ asymptotically.  Then the physical metric in these coordinates is essentially that of Bondi and Sachs.\footnote{Bondi and Sachs actually used a luminosity distance rather than an affine parameter, but the latter is usually more convenient mathematically.}  
The original analysis was for the strict vacuum case, but the present one would accommodate the stress--energies usually accepted for radiation fields (like electromagnetism).

Associated with the Bondi frame at $\scrif$ is a null tetrad, where $n^a$ is tangent to the null generators of $\scrif$, the vector $m^a$ is a antiholomorphic tangent to the $u=\const$ cuts, and $l^a$ is a null orthogonal to the cuts.  (The only non-zero inner products are $l_an^a=1$, $m_a{\overline m}^a=-1$.)  From these unphysically normalized null vectors an associated physically normalized tetrad can be obtained, by requiring the vectors to be (physically) parallel transported along the generators of the $u=\const$ null hypersurfaces and taking ${\widehat n}^a=n^a$, ${\widehat m}^a=\Omega m^a$, ${\widehat l}^a=\Omega^{2}l^a$.  


The particular fields which enter are some of the Newman--Penrose spin coefficient and curvature quantities, defined with respect to the tetrad adapted to the Bondi frame.  
Of special note are the {\em Bondi shear} $\sigma$, measuring the rate of astigmatic change of the $l^a$ congruence as it passes through $\scrif$, and the components $\Psi_n$ (for $0\leq n\leq 4$) of the Weyl tensor at $\scrif$.
These components correspond to the {\em physical} fall-offs ${\widehat\Psi}_n\sim O(r^{5-n})$ (this is the {\em Sachs peeling property}), so the higher values of $n$ correspond to longer-range effects.
It is $\Psi_4$ and $\Psi_3$ which, at $\scrif$, carry the information of the radiation, and one 
has $\Psi_4=-\ddot{\overline\sigma}$, $\Psi_3 =-\eth\dot{\overline\sigma}$ there.  The quantity $-\dot{\overline\sigma}$ thus is a potential for the radiation; it is the {\em Bondi news}.  (Because of the properties of spin-weight functions, the news is uniquely determined from $\Psi_3$.)
The component $\Psi_2$ contains the lead Newtonian term, and $\Psi_1$ what one usually thinks of as the main contribution to the angular momentum.

I should remark that the notation has been streamlined because the analysis is wholly at $\scrif$; in other papers, expressions like $\sigma^0$, $\Psi_n^0$ are often used for what appear here as the values of $\sigma$, $\Psi_n$ at $\scrif$.


%

\end{document}